%% file: main.tex
\documentclass[11pt]{article}

\include{init}

\title{Bayesian Approach to Inverse Problems: an Application to NNPDF Closure Testing}
\author[a]{Luigi Del Debbio} 
\author[b,c]{Tommaso Giani} 
\author[a]{Michael Wilson}
\affil[a]{Higgs Centre for Theoretical Physics, School of Physics and Astronomy,
Peter~Guthrie~Tait~Road, Edinburgh EH9 3 FD, United Kingdom.}
\affil[b]{Department of Physics and Astronomy, Vrije Universiteit, NL 1081 HV Amsterdam}
\affil[c]{Nikhef Theory Group, Science Park 105, 1098 XG Amsterdam, The Netherlands}

\date{}
\makeindex

\begin{document}

\maketitle

\begin{abstract}
    We discuss the Bayesian approach to the solution of inverse problems and
    apply the formalism to analyse the closure tests performed by the NNPDF
    collaboration. Starting from a comparison with the approach that is
    currently used for the determination of parton distributions (PDFs) by the
    NNPDF collaboration, we discuss some analytical results that can be obtained
    for linear problems and use these results as a guidance for the more
    complicated non-linear problems. We show that, in the case of Gaussian
    distributions, the posterior probability density of the parametrized PDFs is
    fully determined by the results of the NNPDF fitting procedure. In the
    particular case that we consider, the fitting procedure and the Bayesian
    analysis yield exactly the same result. Building on the insight that we
    obtain from the analytical results, we introduce new estimators to assess
    the statistical faithfulness of the fit results in closure tests. These
    estimators are defined in data space, and can be studied analytically using
    the Bayesian formalism in a linear model in order to clarify their meaning.
    Finally we present numerical results from a number of closure tests
    performed with current NNPDF methodologies. These further tests allow us to
    validate the NNPDF4.0 methodology and provide a quantitative comparison of the
    NNPDF4.0 and NNPDF3.1 methodologies. As PDFs determinations move into precision
    territory, the need for a careful validation of the methodology becomes
    increasingly important: the error bar has become the focal point of
    contemporary PDFs determinations. In this perspective, theoretical
    assumptions and other sources of error are best formulated and analysed in
    the Bayesian framework, which provides an ideal language to address the
    precision and the accuracy of current fits. 

\end{abstract}

\include{introduction}
\include{inverse}
\include{closure_test}
\include{estimators}
\include{results}
\include{summary}
\appendix
\include{gaussian}
\include{app_datasets}
\include{app_deltachi2}
\bibliographystyle{UTPstyle}
\bibliography{main}

\end{document}

%% file: init.tex
\usepackage{amsmath}
\usepackage{amsfonts}
\usepackage{amssymb}
\usepackage{authblk}
\usepackage{dsfont}
\usepackage{pifont}
\usepackage{booktabs}
\usepackage{tabularx}
\usepackage{siunitx}
\usepackage{graphicx}
\usepackage{epstopdf}
\usepackage{epsfig}
\usepackage{framed}
\usepackage{makeidx}
\usepackage{simplewick}
\usepackage{hyperref}
\usepackage{placeins}
\usepackage{bbold}
\usepackage[font=small,labelfont=bf]{caption}
\usepackage{tikz}
\usetikzlibrary{positioning}
\usetikzlibrary{arrows}
\usetikzlibrary{colorbrewer}

\DeclareSIUnit{\barn}{b}

\newcommand{\cov}{C}
\newcommand{\posterior}[1]{\tilde{#1}}
\newcommand{\real}{\mathbb{R}}

\newcommand{\fwdmapop}{G}

\newcommand{\fwdobsop}{\mathcal{\fwdmapop}}
\newcommand{\nmodel}{N_{\rm model}}
\newcommand{\modelspace}{X}
\newcommand{\modelvec}{u}
\newcommand{\modelpriorcent}{\modelvec_0}
\newcommand{\modelpriorcov}{\cov_X}
\newcommand{\modelpostcent}{\posterior{\modelvec}}
\newcommand{\modelpostcov}{\posterior{\cov}_X}
\newcommand{\ndata}{N_{\rm data}}
\newcommand{\obs}{y}
\newcommand{\obspriorcent}{\obs_0}
\newcommand{\obspriorcov}{\cov_{Y}}
\newcommand{\obsnoise}{\eta}
\newcommand{\obspostcent}{\posterior{\obs}}
\newcommand{\obspostcov}{\posterior{\cov}_Y}
\newcommand{\linmap}{\fwdobsop}

\newcommand{\nlaw}{N_{\rm law}}

\newcommand{\law}{f}
\newcommand{\pseudodat}{\mu}
\newcommand{\noise}{\epsilon}
\newcommand{\repind}{(k)}
\newcommand{\modelvecrep}{\modelvec_*^{\repind}}
\newcommand{\likelihood}{\mathcal{L}}
\newcommand{\repchis}{{\chi^2}^{\repind}}

\newcommand{\lawmodel}{w}
\newcommand{\utrue}{u_\mathrm{true}}
\newcommand{\uest}{u_\mathrm{est}}

\newcommand{\testset}[1]{ {{#1}^{\prime}} }
\newcommand{\emodel}[1]{ \mathbf{E}_{\{ \modelvec_* \}} \left[ #1 \right] }
\newcommand{\eout}{\mathcal{E}^{\rm out}}

\newcommand{\nfits}{N_{\rm fits}}
\newcommand{\nreps}{N_{\rm replicas}}

\newcommand{\bias}{{\rm bias}}
\newcommand{\var}{{\rm variance}}
\newcommand{\covrep}{\testset{\cov}^{(\rm replica)}}
\newcommand{\covcent}{\testset{\cov}^{(\rm central)}}
\newcommand{\biasvarratio}{\mathcal{R}_{bv}}

\newcommand{\xisigdat}[1]{\xi^{(\rm data)}_{#1 \testset{\sigma}}}
\newcommand{\xisigdati}[1]{\xi^{(\rm data)}_{#1 {\testset{\sigma}_i}}}
\newcommand{\modelstd}{\hat{\sigma}}
\newcommand{\erf}{{\rm erf}}

\newcommand{\ie}{{\it i.e.}}
\newcommand{\eg}{{\it e.g.}}
\newcommand{\viz}{{\it viz.}}

\newcommand{\ein}{\mathcal{E}^{\rm in}}
\newcommand{\deltachi}{\Delta_{\chi^2}}
\newcommand{\noisecross}{{\rm noise \, cross \, term}}

\newcommand{\shift}{\obsnoise}

\graphicspath{{./figures/}}

%% file: introduction.tex
\section{Introduction}
\label{sec:Intro}

Inverse problems are the typical example of inference where a model is sought
starting from a finite-dimensional set of experimental observations. These
problems are notoriously difficult, and often require trying to guess a
continuous function, \ie\ an element of an infinite dimensional space, from a
finite amount of data. As emphasised by Hadamard a long time ago, it is easy to
end up in a situation where we deal with ill-posed problems, in the sense that
the solution may not exist, may not be unique, or may be unstable under small
variations of the input data. The determination of parton distributions from
experimental data, or the reconstruction of spectral densities from lattice QCD
simulations, are just two examples where these problems arise in particle
physics. In all these cases, finding a robust solution becomes very challenging,
if not impossible, making these questions all the more urgent, especially at a
time when precision studies are the ultimate challenge in order to highlight
divergences from the Standard Model. 

A Bayesian approach provides an apter tool for addressing inverse problems.
Starting from a prior distribution for the model, which encodes our theoretical
knowledge or prejudice, basic statistical tools allow us to determine the
posterior distribution of the solution, {\it after taking into account a set of
experimental observations.} The prior and posterior probabilities encode our
knowledge about the model before and after incorporating the experimental
results. There are multiple advantages to this formulation: the inverse problem
is well defined, and the prior distribution ensures that all the assumptions are
explicit, while regulating the likelihood distribution. 

Note also that probability measures can be defined in infinite-dimensional
spaces. In cases where we are looking to determine a continuous function, the
Bayesian approach provides, at least in principle, a formulation of the problem
directly in terms of the posterior probability measure of the model function. It
is often convenient for practical reasons to define a parametrization of the
model function in terms of a finite, albeit large, number of parameters and
reduce the problem to a finite dimensional one. The price to pay for this
simplification is the introduction of some bias, which needs to be understood
and possibly quantified. An expressive parametrization clearly helps in this
case. 

A Bayesian approach to inverse problems has been actively developed by
mathematicians for a long time, and this development has dramatically
accelerated in the last decade. In this paper we aim at summarising the existing
framework and adapt it to analyse the fits of parton distribution functions
obtained by the NNPDF collaboration. We review the Bayesian formalism in
Sec.~\ref{sec:inverse-problems}, where we define the notation that will be used
in the rest of the
paper. We report some known examples for illustrative purposes. Even though
these are well-known results, we find it useful to summarise them in the context
of our specific problem. In Sec.~\ref{sec:closure-test}, we try to connect the
Bayesian approach
with the NNPDF fits based on a Monte Carlo methodology, where the distribution
of the PDFs is encoded in a set of fits to artificially generated data, called
replicas. We can anticipate here that, under the hypotheses that the data are
Gaussian and the model is linear, the NNPDF procedure does characterise
completely the posterior proabillity density. When the model is non-linear two
modifications need to be taken into account. First of all the analytical
calculation that we present in Sec.~\ref{sec:BayesianInverse} is no longer
possible, and one needs to rely on the fact that a linearization of the model
yields an accurate prediction. Even though linear models are known to provide
good approximations, the systematic errors introduced by this approximation are
not easy to quantify. There is also a more subtle effect that needs to be taken
into account. When working with linear models, the minimization procedure is
know to have a unique minimum, which can be computed exactly. Non-linear models
can be plagued by multiple minima, and more importantly by inefficiencies of the
algorithm in finding them. While it would be foolish to ignore the limitations
of the analytical calculation, it is nonetheless very useful to have an explicit
result as a template to guide the analysis of our numerical investigations. 

In the NNPDF fitting framework, the posterior probability of the parton
distribution functions, is encoded in an ensemble of fits to replicas of the
data, where the data replicas have been generated in order to reproduce the
fluctuations described by the experimental central values and uncertainties.
This bootstrap procedure propagates the data fluctuations in the space of fitted
PDFs. A successful fit should yield robust confidence intervals for observables,
in particular for those that are more relevant for phenomenology.

The idea of {\em closure tests} is to test this procedure in a fit to artificial
data that have been generated from a known set of input PDFs. In this case the
underlying law is known and we can check how the posterior distribution compares
to the underlying law. This is the basis of a closure test, which is summarised
at the end of Sec.~\ref{sec:closure-test}. Closure tests have already been used
to test to validity
of previous iterations of the NNPDF methodology. Here we aim to refine some of
the pre-existing closure test estimators and with the help of fast fitting
methodology perform a more extensive study of how faithful our uncertainties
are. For this purpose we introduce in Sec.~\ref{sec:ClosureEstimators} new
estimators that allow us to
quantify the quality of our fits. These estimators are defined in the space of
data and need to be understood as stochastic variables that are characterised by
a probability density. Where possible, we use the Bayesian formalism in order to
compute analytically these probability densities and compare with numerical
results. In order to perform analytical calculations we often need to make some
simplifying assumptions. While it is incorrect to use the analytical results for
making quantitative predictions for the results of the realistic case of a
non-linear fit, the analytical results provide a guide to the interpretation of
the observed patterns. 

The results of our numerical studies are summarised in
Sec.~\ref{sec:numerical-results}, where we also
compare the NNPDF4.0 methodology introduced in the latest NNPDF
fit~\cite{NNPDF40} to the methodology used by the collaboration in previous
fits. 
 
The Bayesian formalism, by providing posterior probability distributions, paves
the way to explore a number of issues. We highlight in our conclusions some
possible questions that we defer to future studies. 

%% file: inverse.tex
\section{Inverse Problems}
\label{sec:inverse-problems}

The problem of determining PDFs from a set of experimental data falls under the
general category of {\em inverse problems}, \ie\ the problem of finding the
input to a given model knowing a set of observations, which are often finite and
noisy. In this section we are going to review the Bayesian formulation of
inverse problems. It is impossible to do justice to this vast subject here.
Instead we try to emphasise the aspects that are relevant for quantifying
uncertainties on PDF determinations. 

\subsection{Statement of the problem}
\label{sec:BayesianInverse}

The space of inputs is denoted by $\modelspace$, while $R$ denotes the space of
responses. The model is specified by a {\em forward map}
\begin{align}
  \label{eq:ForwardMap}
  \fwdmapop : ~& \modelspace \to R \nonumber \\
      & \modelvec \mapsto r=\fwdmapop(\modelvec) \, ,
\end{align}
which associates a response $r \in R$ to the input $\modelvec \in \modelspace$,
where we assume that $\modelspace$ and $R$ are Banach spaces.~\footnote{Banach
spaces are complete normed vector spaces. We do not need to get into a more
detailed discussion here, but it is important to note that working in Banach
spaces allows us to generalise the results to infinite-dimensional spaces of
functions.} As an example we can think of $\modelvec$ as being a PDF, \ie\ a
function defined on the interval $[0,1]$, and $r$ a DIS structure function. The
structure function is related to the PDF by a factorization formula involving
perturbative coefficient functions: 
\begin{align}
  \label{eq:DISExample}
  r(x,Q^2) = \int_x^1 \frac{dz}{z}\, C(z,Q^2) \modelvec(x/z,Q^2)\, .
\end{align}
Note that in this example the forward map maps one real function into another
real function. In this case the space of models $X$ is the space of continuous
functions defined on the interval $[0,1]$, which satisfy integrability
conditions. Even though this is an infinite-dimensional space, it is possible to
define a probability measure on such a space and construct a Bayesian solution
to the inverse problem. In current determinations of PDFs, the functional form
of the PDF is dictated by some kind of parametrization, with different
parametrizations being used by different collaborations. In all cases, the space
$X$ is a finite-dimensional space, $\real^{\nmodel}$, where $\nmodel$ is
the number of parameters. In the case of the NNPDF fits discussed below, the
weights of the neural networks are the parameters that determine the functional
form of the PDFs. Alternatively, one could think of a finite-dimensional
representation defined by the value of the PDF at selected values of $x$, \ie\
$u_i=u(x_i)$ for $i=1,\ldots,\nmodel$. Depending on the context we will denote
by $u$ either the function $u(x)$, or the vector of real parameters that are
used to determine the function $u(x)$. Disambiguation should hopefully be
straightforward. 

Experiments will not have access to the full function $r$ but only to a subset
of $\ndata$ observations. In order to have a formal mathematical expression that
takes into account the fact that we have a finite number of measurements, we
introduce an {\em observation operator}
\begin{align}
  O : ~& R \to Y \nonumber \\
       & r \mapsto \obs \, ,
\end{align}
where $\obs \in Y$ is a vector in a finite-dimensional space $Y$ of experimental
results, \eg\ the value of the structure function for some values of the
kinematic variables $x$ and $Q^2$. In general we will assume that $\obs \in
\real^{\ndata}$, \ie\ we have a finite number $\ndata$ of real experimental
values. The quantity of interest is the composed operator
\begin{align}
  \fwdobsop : ~& \modelspace \to \real^{\ndata} \nonumber \\
                 & \fwdobsop = O \circ G\, ,
\end{align}
which maps the input $\modelvec$ to the set of data. Taking into account the
fact that experimental data are subject to noise, we can write
\begin{align}
  \label{eq:NoisyInverseProblem}
  \obs = \fwdobsop(\modelvec) + \obsnoise\, ,
\end{align}
where $\obsnoise$ is a random variable defined over $\real^{\ndata}$ with
probability density $\rho(\obsnoise)$. We will refer to $\obsnoise$ as the {\em
observational noise}. In this setting, the inverse problem becomes finding
$\modelvec$ given $\obs$. It is often the case that inverse problems are
ill-defined in the sense that the solution may not exist, may not be unique, or
may be unstable under small variations of the problem. 

In solving the inverse problem, we are going to adopt a Bayesian point of view,
as summarised \eg\ in Ref.~\cite{Stuart:2010}: our prior knowledge about
$\modelvec$ is encoded in a prior probability measure $\mu_X^0$, where the
suffix $X$ indicates that the measure is defined in the space of models, and the
suffix 0 refers to the fact that this is a prior distribution. We use Bayes'
theorem to compute the posterior probability measure of $\modelvec$ given the
data $\obs$, which we denote as $\mu_X^\fwdobsop$. When the probability measure
can be described by a probability density, we denote the probability densities
associated to $\mu_X^0$ and $\mu_X^\fwdobsop$, by $\pi_X^0$ and
$\pi_X^\fwdobsop$ respectively. Then, using Eq.~(\ref{eq:NoisyInverseProblem}),
we can write the data likelihood, \ie\ the probability density of $\obs$ given
$\modelvec$,
\begin{align}
  \label{eq:YGivenUProbDensity}
  \pi_Y(\obs|\modelvec) = \rho(\obs-\fwdobsop(\modelvec))\, ,
\end{align}
and Bayes' theorem yields
\begin{align}
  \label{eq:BayesThmInversePosterior}
  \pi_X^\fwdobsop(\modelvec) = \pi_X(\modelvec|\obs) \propto 
  \pi_X^0(\modelvec)
  \rho(\obs-\fwdobsop(\modelvec))\, .
\end{align}

Even though the concepts that we have introduced so far should sound familiar,
it is worthwhile clarifying some ideas and present an explicit case where all
the probability densities are carefully defined. This is best exemplified by
considering the case where both the observational noise and the model prior are
Gaussian. We assume that we are given a set of central values $\obspriorcent \in
\real^{\ndata}$ and their covariance matrix $\obspriorcov$. Then the {\em prior}
probability density of the observable $\obs$ is 
\begin{equation}
  \label{eq:PriorData}
  \pi_{Y}^0(\obs|\obspriorcent,\obspriorcov) \propto \exp\left(
    -\frac12 \left| \obs - \obspriorcent \right|_{\obspriorcov}^2
    \right)\, ,
\end{equation}
where, similarly to the convention used above, the suffix $Y$ emphasises the
fact that this is a probability density in data space, and the notation
explicitly reminds us that this is the probability density given the central
values $\obspriorcent$ (and the covariance matrix $\obspriorcov$). Similarly we
can choose a Gaussian distribution for the prior distribution of the input
model, characterized by a central value $\modelpriorcent$ and a covariance
$\modelpriorcov$:
\begin{align}
  \label{eq:PiZeroGauss}
  \pi_{X}^0(\modelvec|\modelpriorcent,\modelpriorcov)  
  &\propto \exp\left(
              -\frac12 \left| \modelvec - \modelpriorcent \right|_{\modelpriorcov}^2
              \right)\, .
\end{align}
Following the convention above, we use a suffix $X$ here to remind the reader
that we are looking at a probability density in the space of models. Note that
in the expressions above we used the norms in $\modelspace$ and $\real^{\ndata}$
respectively, and introduced the short-hand notation
\begin{align}
  \left|a\right|_M^2 = \left| M^{-1/2} a\right|^2\, ,
\end{align}
where $a$ denotes a generic element of $\modelspace$, $R$ or $\real^{\ndata}$.
For the case where $a \in \real^{\ndata}$, we use the Euclidean norm and
\begin{align}
  \left| a \right|_M^2 = \sum_{i,j} a_i M^{-1}_{ij} a_j\, ,
\end{align}
where the indices $i,j$ run from 1 to $\ndata$, which eventually yields the
usual expression for the $\chi^2$ of correlated data.   
Up to this point data and models are completely independent, and the joint
distribution is simply the product of $\pi_{Y}^0$ and $\pi_{X}^0$. 

The forward map induces a correlation between the input model and the
observables, so we introduce a probability density $\theta$ that describes these
correlations due to the underlying theory,  
\begin{equation}
  \label{eq:ThetaCorr}
  \theta(\obs,\modelvec|\fwdobsop) = \delta\left(\obs - \fwdobsop(\modelvec)\right)\, ,
\end{equation}
where the Dirac delta corresponds to the case where there are no theoretical
uncertainities. Theoretical uncertainties can be introduced by broadening the
distribution of $\obs$ away from the exact prediction of the forward map, \eg\
using a Gaussian with covariance $C_T$,
\begin{equation}
  \label{eq:TheoryErrors}
  \theta(\obs,\modelvec|\fwdobsop) = \exp\left(
    -\frac12 
    \left| \obs - \fwdobsop(\modelvec)
    \right|_{C_T}^2\right)\, .
\end{equation}
In the context of PDF fitting a similar recipe to take into account theoretical
errors has recently been advocated in
Refs.~\cite{NNPDF:2019vjt,AbdulKhalek:2019ihb}. Note that there are no rigorous
arguments favouring the assumption that theoretical errors are normally
distributed; it is nonetheless a useful working assumption, and a definite
improvement compared to ignoring the theoretical errors altogether. The net
effect of the theory errors is a redefinition of the covariance of the data,
which has no major impact in our discussion, and therefore will be ignored.
Taking the correlation $\theta(\obs,\modelvec|\fwdobsop)$ into account, the
joint distribution of $\obs$ and $\modelvec$ is
\begin{align}
  \label{eq:JointYAndU}
  \pi^\fwdobsop(\obs,\modelvec|\obspriorcent,\obspriorcov,\modelpriorcent,\modelpriorcov) 
  \propto 
  \pi_{X}^0(\modelvec|\modelpriorcent, \modelpriorcov) 
  \pi_{Y}^0(\obs|\obspriorcent,\obspriorcov) 
  \theta(\obs,\modelvec|\fwdobsop)\, .
\end{align}
We can now marginalize with respect to \obs, neglecting theory errors, 
\begin{align}
  \label{eq:MarginOne}
  \pi^\fwdobsop_{X}(\modelvec|\obspriorcent,\obspriorcov,\modelpriorcent,\modelpriorcov) 
  &\propto \int dy\, 
    \pi_{X}^0(\modelvec|\modelpriorcent,\modelpriorcov) 
    \pi_{Y}^0(\obs|\obspriorcent,\obspriorcov) 
    \theta(\obs,\modelvec|\fwdobsop) \\
  & \propto \pi_{X}^0(\modelvec|\modelpriorcent,\modelpriorcov)  
    \int dy\, 
    \pi_{Y}^0(\obs|\obspriorcent,\obspriorcov) 
    \delta\left(\obs-\fwdobsop(\modelvec)\right) \\
  & \propto 
    \pi_{X}^0(\modelvec|\modelpriorcent,\modelpriorcov) \,
    \pi_{Y}^0(\fwdobsop(\modelvec)|\obspriorcent,\obspriorcov)\, .
\end{align}
We see that we have recovered Eq.~\ref{eq:BayesThmInversePosterior}. The
log-likelihood in the Gaussian case is simply the $\chi^2$ of the data,
$\obspriorcent$, to the theory prediction, $\fwdobsop(\modelvec)$:
\begin{equation}
  \label{eq:LikelyChiSq}
  -\log\pi_{Y}^0(\fwdobsop(\modelvec)|\obspriorcent,\obspriorcov) =  
    \frac12 \sum_{i,j=1}^{\ndata}
      \left(\fwdobsop(\modelvec) - \obspriorcent \right)_i
      \left(\obspriorcov^{-1}\right)_{ij}
      \left(\fwdobsop(\modelvec) - \obspriorcent \right)_j
    \, .
\end{equation}
In the notation of Eq.~\ref{eq:BayesThmInversePosterior}
\begin{equation}
  \label{eq:IdentifyRho}
  \pi_{Y}^0(\fwdobsop(\modelvec)|\obspriorcent,\obspriorcov) = \rho\left(
    \fwdobsop(\modelvec) - \obspriorcent
  \right)\, ,
\end{equation}
where in this case 
\begin{align}
  \label{eq:RhoGauss}
  \rho(\obsnoise) &\propto \exp\left(
               -\frac12 \left|\obsnoise\right|_{\obspriorcov}^2
               \right)\, .
\end{align}
The probability density $\pi^\fwdobsop_{X}(\modelvec|\obspriorcent,
\obspriorcov,\modelpriorcent,\modelpriorcov)$ was called
$\pi_{X}^\fwdobsop(\modelvec)$ in Eq.~\ref{eq:BayesThmInversePosterior}, where
the suffix $\fwdobsop$ is a short-hand to denote the posterior probability in
model space, taking into account all the conditional variables. Hence, for the
Gaussian case, the result from Bayes' theorem reduces to
\begin{align}
  \label{eq:PosteriorModel}
  \pi_{X}^\fwdobsop(\modelvec) 
  &\propto 
  \exp\left[
    -\frac12 \left| \obspriorcent - \fwdobsop(\modelvec) \right|_{\obspriorcov} ^2
    -\frac12 \left| \modelvec - \modelpriorcent \right|_{\modelpriorcov}^2
  \right] \\ 
  &\propto
  \exp\left[
    - S(\modelvec)
  \right]\, .
\end{align}
Note that in the argument of the likelihood function we have the central values
of the data points $\obspriorcent$ as given by the experiments.
Eq.~(\ref{eq:PosteriorModel}) is the Bayesian answer to the inverse problem, our
knowledge of the model $\modelvec$ is encoded in the probability measure
$\mu_{X}^\fwdobsop$, which is fully specified by the density
$\pi_{X}^\fwdobsop$. There are several ways to characterise a probability
distribution, a task that becomes increasingly difficult in high-dimensional
spaces. As discussed later in this study, the NNPDF approach is focused on the
determination of the {\em Maximum A Posteriori (MAP)} estimator, \ie\ the
element $u_* \in \modelspace$ that maximises $\pi_{X}^\fwdobsop(\modelvec)$:
\begin{align}
  \label{eq:MAP}
  u_* = \arg\min_{\modelvec \in \modelspace} 
  \left(
    \frac12 \left| \obspriorcent - \fwdobsop(\modelvec) \right|_{\obspriorcov}^2
    + \frac12 \left| \modelvec - \modelpriorcent \right|_{\modelpriorcov}^2
  \right)\, .
\end{align}
For every instance of the data $\obspriorcent$, the MAP estimator is computed by
minimising a regulated $\chi^2$, where the regularization is determined by the
prior that is assumed on the model $u$. We will refer to this procedure as the
{\em classical fit} of experimental data to a model. Note that in the Bayesian
approach, the regulator appears naturally after having specified carefully all
the assumptions that enter in the prior. In this specific example the regulator
arises from the Gaussian prior for the model input $\modelvec$, which is
normally distributed around a solution $\modelpriorcent$. The MAP estimator
provides the explicit connection between the Bayesian approach and the classical
fit.

\subsection{Comparison with classical fitting}
\label{sec:comp-class-fit}

Analytical results make the connection between the two approaches more
quantitative, and therefore more transparent. We are going to summarise these
results here without proofs, referring the reader to the mathematical literature
for the missing details. Working in the finite-dimensional case, we assume 
\begin{align*}
  \modelvec &\in \real^{\nmodel} \, ,\\
  \obs &\in \real^{\ndata}\, ,
\end{align*}
and we are going to review in detail two examples from Ref.~\cite{Stuart:2010},
which illustrate the role of the priors in the Bayesian setting. We report here
the results in Ref.~\cite{Stuart:2010} because their simplicity provides a
striking example of the role of priors, which is sometimes underestimated. It is
particularly useful to distinguish the case of an underdetermined system from
the case of an overdetermined one. 

\paragraph{Underdetermined system}
The first case that we are going to analyse is the case of a linear system that
is underdetermined by the data. The linear model is completely specified by a
vector of coefficients $g\in \real^{\nmodel}$, 
\begin{equation}
  \label{eq:LinSyst}
  \mathcal{G}(u) = \left(g^T u\right)\, .
\end{equation}
Assuming that we have only one datapoint, \ie\ $\ndata=1$, 
\begin{equation}
  \label{eq:LinearModelEx}
  \obs = (g^T \modelvec) + \obsnoise\, ,
\end{equation}
where $\obsnoise \sim \mathcal{N}(0,\gamma^2)$ is one Gaussian number, whose
probability density is centred at $0$ and has variance $\gamma^2$. 
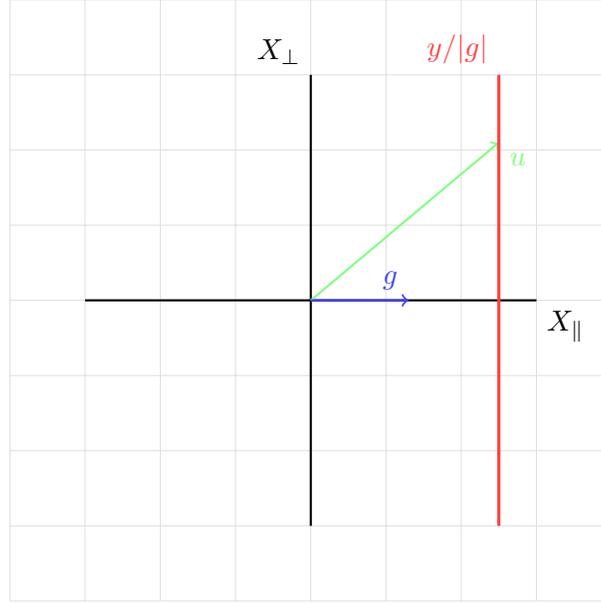
\begin{figure}[h!]
  \centering
  \begin{tikzpicture}
      \draw[step=1cm,gray!25!,very thin] (-4,-4) grid (4,4);
      \draw[thick] (-3,0) -- (3,0) node[anchor=north west] {$X_\parallel$};
      \draw[thick] (0,-3) -- (0,3) node[anchor=south east] {$X_\perp$};
      \draw[green!50,thick,->] (0,0) -- (2.5,2.1) node[anchor=north west] {$u$};
      \draw[blue!75,thick,->] (0,0) -- (1.3, 0) node[anchor=south east] {$g$};
      \draw[red!75,very thick] (2.5,-3) -- (2.5, 3) node[anchor=south east] 
        {$y/|g|$};
  \end{tikzpicture}
  \caption{A linear model in a two-dimensional space is constrained by a single data point $y$. Once the vector $g$ is given, the model is fully specified by the vector $u$, \viz\ $y=g^T u$. All models $u$ along the vertical red line reproduce exactly the data point. In the absence of prior knowledge, every model along that one-dimensional subspace is a legitimate solution of the inverse problem, which is clearly underdetermined.}
  \label{fig:2dexample}
\end{figure}
For simplicity we are going to assume that the prior on $\modelvec$ is also a
multi-dimensional Gaussian, centred at $0$ with covariance matrix
$\modelpriorcov$. In this case the posterior distribution can be written as
\begin{equation}
  \label{eq:GaussPostExplicit}
    \pi_{X}^\fwdobsop(\modelvec) 
    \propto \exp \left[
      -\frac{1}{2\gamma^2} \left|\obs - (g^T \modelvec) \right|^2 - \frac12 \left|\modelvec\right|_{\modelpriorcov}^2 
    \right]\, ,
\end{equation}
which is still a Gaussian distribution for $\modelvec$. The mean and covariance
are respectively
\begin{align}
  m &= \frac{(\modelpriorcov g) \obs}{\gamma^2 + (g^T \modelpriorcov g)}\, , \\
  \Sigma &= \modelpriorcov - 
  \frac{(\modelpriorcov g) (\modelpriorcov g)^T}{\gamma^2 + (g^T \modelpriorcov g)}\, .
\end{align}
Because the argument of the exponential is a quadratic form, the mean of the
distribution coincides with the MAP estimator. Hence in this case a fit that
minimises the $\chi^2$ of the data to the theory prediction yields exactly the
mean of the posterior distribution. It is instructive to look at these
quantities in the limit of infinitely precise data, \ie\ in the limit $\gamma\to
0$:
\begin{align}
  m_\star &= 
  \lim_{\gamma\to 0} m
  = \frac{(\modelpriorcov g) \obs}{(g^T \modelpriorcov g)}\, , \\
  \Sigma_\star &= 
  \lim_{\gamma\to 0} \Sigma 
  = \modelpriorcov - 
  \frac{(\modelpriorcov g) (\modelpriorcov g)^T}{(g^T \modelpriorcov g)}\, .
\end{align}
These values satisfy
\begin{align}
  (g^T m_\star) = \obs \, , \\
  (\Sigma_\star g) = 0 \, ,
\end{align}
which shows that the mean of the distribution is such that the data point is
exactly reproduced by the model, and that the uncertainty in the direction
defined by $g$ vanishes. It should be noted that the uncertainty in directions
perpendicular to $g$ does not vanish and is determined by a combination of the
prior and the model, \viz\ $\modelpriorcov$ and $g$ in our example. This is a
particular example of a more general feature: for underdetermined systems the
information from the prior still shapes the probability distribution of the
solution even in the limit of vanishing statistical noise.  

It is interesting to analyse what happens when the prior on the model is
removed. For this purpose we can start from a covariance in model space that is
proportional to the identity matrix, 
\begin{equation}
  \label{eq:DiagCovSigma}
  \modelpriorcov = \sigma \mathbb{1}\, ,
\end{equation}
and take the limit $\sigma \to \infty$. In this limit
\begin{equation}
  \label{eq:LargeSigmaLimit}
  m_\star = y \frac{g}{(g^T g)}\, ,
\end{equation}
while the posterior covariance becomes
\begin{equation}
  \label{eq:ExplcitCov}
  \Sigma_\star= \sigma \left(
    \mathbb{1} - \frac{g g^T}{(g^T g)}
  \right)\, .
\end{equation}
This covariance is the projector on the subspace orthogonal to $g$ multiplied by
$\sigma$. In the simple two-dimensional example depicted in
Fig.~\ref{fig:2dexample}, choosing the direction of $g$ as one of the basis
vector, we obtain
\begin{equation}
  \label{eq:ExplCovTwo}
  \Sigma_\star = 
  \begin{pmatrix}
    0 & \\
    & \sigma
  \end{pmatrix}\, ,
\end{equation}
which shows that the error in the $g$ direction vanishes, while the error in the
orthogonal subspace diverges as we remove the prior. 

\paragraph{Overdetermined system}
We are now going to consider an example of an overdetermined system and discuss
again the case of small observational noise. We consider $\ndata\geq 2$ and
$\nmodel=1$, with a linear forward map such that
\begin{equation}
 \label{eq:OverDetForwMap}
 \obs = g \modelvec  + \obsnoise\, ,
\end{equation} 
where $\obsnoise$ is an $\ndata$-dimensional Gaussian variable with a diagonal
covariance $\gamma^2 \mathbb{1}$, and $\mathbb{1}$ denotes the identity matrix.
For simplicity we are going to assume a Gaussian prior with unit variance for
$\modelvec$, which yields for the posterior distribution:
\begin{equation}
  \label{eq:OverDetPost}
  \pi_{X}^\fwdobsop(\modelvec) 
    \propto 
    \exp\left(
      -\frac{1}{2\gamma^2} \left| \obs - g \modelvec \right|^2
      -\frac12 \modelvec^2
    \right)\, .
\end{equation} 
The posterior is Gaussian and we can easily compute its mean and variance: 
\begin{align}
  m &= \frac{(g^T \obs)}{\gamma^2 + |g|^2} \, , \\
  \sigma^2 &=
    \frac{\gamma^2}{\gamma^2 + |g|^2}\, .
\end{align}
In this case, in the limit of vanishing observational noise, we obtain
\begin{align}
  m_\star &= \frac{(g^T \obs)}{|g|^2} \, ,\\
  \sigma_\star^2 &= 0\, .
\end{align}
The mean is given by the weighted average of the datapoints, which is also the
solution of the $\chi^2$ minimization
\begin{equation}
  m_\star = \arg\min_{\modelvec\in\real} \left|\obs - g \modelvec\right|^2\, .
\end{equation}
Note that in this case the variance $\sigma_\star$ vanishes independently of the
prior. In the limit of small noise, the distribution tends to a Dirac delta
around the value of the MAP estimator.  

\subsection{Linear Problems}
\label{sec:LinProbs}

Linear problems in finite-dimensional spaces are characterized by a simple
forward law, 
\begin{equation}
  \label{eq:MatrixG}
  \obs = \linmap \modelvec\, ,
\end{equation}
where $\linmap$ is a matrix. In this framework one can readily  derive
analytical solutions that are useful to understand the main features of the
Bayesian approach. Assuming that the priors are Gaussian again, the cost
function $S(\modelvec)$ is a quadratic function of $\modelvec$,
\begin{align}
  \label{eq:CostLinGauss}
  S(\modelvec) &= 
  \left(\linmap \modelvec - \obspriorcent \right)^T \obspriorcov^{-1} 
  \left(\linmap \modelvec - \obspriorcent \right) + 
  \left( \modelvec - \modelpriorcent \right)^T \modelpriorcov^{-1} \left(\modelvec - \modelpriorcent \right) \\
  &= 
  \left(\modelvec - \modelpostcent\right) \modelpostcov^{-1}
  \left(\modelvec - \modelpostcent\right) + \mathrm{const}\, ,
\end{align} 
where
\begin{align}
  \label{eq:PostParamsCov}
  \modelpostcov^{-1} &= 
  \left(
    \linmap^T \obspriorcov^{-1} \linmap + \modelpriorcov^{-1}
  \right)\, , \\
  \label{eq:PostParamsMean}
  \modelpostcent &=
  \modelpostcov  \left(
    \linmap^T \obspriorcov^{-1} \obspriorcent + \modelpriorcov^{-1} \modelpriorcent
  \right)\, .
\end{align}
The case where we have no prior information on the model is recovered by taking
the limit $\modelpriorcov^{-1} \to 0$, which yields
\begin{align}
  \label{eq:NoPriorLinModel}
  \modelpostcov^{-1} &= 
  \left(
    \linmap^T \obspriorcov^{-1} \linmap
  \right)\, , \\
  \modelpostcent &=
  \modelpostcov  \left(
    \linmap^T \obspriorcov^{-1} \obspriorcent 
  \right)\, . \label{eq:NoPriorLinModelCov}
\end{align}
The action of $\obspriorcov^{-1}$ on the vector of observed data $\obspriorcent$
is best visualised using a spectral decomposition
\begin{equation}
  \label{eq:CDSpecDec}
  \obspriorcov^{-1} = \sum_n \frac{1}{\sigma_n^2} P_n\, ,
\end{equation}
where $P_n$ denotes the projector on the $n$-th eigenvector of $\obspriorcov$,
and $\sigma_n^2$ is the corresponding eigenvalue. The action of
$\obspriorcov^{-1}$ is to perform a weighted average of the components of
$\obspriorcent$ in the directions of the eigenvectors of $\obspriorcov$.

An explicit expression for the posterior distribution of data can be obtained
from the joint distribution by marginalising over the model input $\modelvec$:
\begin{align}
  \label{eq:PosteriorDataSpace}
  \pi_{Y}^\fwdobsop(\obs|\obspriorcent,\obspriorcov,\modelpriorcent,\modelpriorcov)
  &= \int du\, 
  \pi^\fwdobsop(\obs,\modelvec|\obspriorcent,\obspriorcov,\modelpriorcent,\modelpriorcov) \\
  &\propto \exp\left(
    -\frac12 \left(\obs - \obspostcent\right)^T \obspostcov^{-1}
    \left(\obs - \obspostcent\right)
  \right)\, ,
\end{align}
where
\begin{align}
  \label{eq:PosteriorDataParamsMean}
  \obspostcent &= \linmap \modelpostcent\, , \\
  \label{eq:PosteriorDataParamsCov}
  \obspostcov &= \linmap \modelpostcov \linmap^T\, .
\end{align}
Note that this is the naive error propagation from the covariance of the model,
$\modelpostcov$, to the space of data. 

\paragraph{Posterior distribution of unseen data}

In real-life cases we are also interested in the posterior distribution of a set
of data that have not been included in the fit. Because different datasets are
described by the same theory, the knowledge of one dataset will inform our
knowledge of the underlying theory -- \ie\ we will determine a posterior
distribution for the model. That new knowledge about the model will then
propagate to any other -- unseen -- set of data, even if the experiments are
completely unrelated. In the Bayesian framework that we have developed, this
situation can be modeled by having two independent sets of data $y$ and $y'$,
for which we have a prior distribution 
\begin{align}
  \label{eq:JointIndepDataPrior}
  \pi_{Y}^0&\left(y,y'|y_0,C_{Y},y'_0,C'_{Y}\right) 
   = \pi_{Y}^0\left(y'|y'_0,C'_{Y}\right) \pi_{Y}^0\left(y|y_0,C_{Y}\right) \\
  & \propto 
  \exp\left[-\frac12 \left(y'-y'_0\right)^T (C'_{Y})^{-1} 
  \left(y'-y'_0\right)\right]\, 
  \exp\left[-\frac12 \left(y-y_0\right)^T (C_{Y})^{-1} 
  \left(y-y_0\right)\right]\, .
\end{align}
Note that the prior distribution is factorised as the product of individual
distributions for $y$ and $y'$ since the datasets are assumed to be independent.
Following the derivation above, we can write the joint distribution for the data
and the model 
\begin{equation}
  \label{eq:JointModelData}
  \pi^\fwdobsop(y,y',u) 
  \propto 
  \pi_{Y}^0(y,y'|y_0,C_{Y},y'_0,C'_{Y}) 
  \pi_{X}^0(u) 
  \delta\left(y - \mathcal{G}u\right)
  \delta\left(y'- \mathcal{G}'u\right)\, .
\end{equation}
Because both sets of data are derived from the same model $u$, the joint
distribution above introduces a correlation between the data sets. The same
model $u$ appears in both delta functions in the equation above. We can now
marginalise with respect to the dataset $y$, 
\begin{equation}
  \label{eq:MarginaliseDatasetY}
  \begin{split}
    \pi(y',u) 
    \propto &
    \exp\left[-\frac12 \left(y'-y'_0\right)^T (C'_{Y})^{-1} 
    \left(y'-y'_0\right)\right]\, 
    \exp\left[-\frac12 \left(u-\tilde{u}\right)^T (\tilde{C}_{X})^{-1} 
    \left(u-\tilde{u}\right)\right] \\
    & \quad \times \delta\left(y'- \mathcal{G}'u\right)\, .
  \end{split}
\end{equation}
where $\tilde{C}_{X}$ and $\tilde{u}$ are given respectively in
Eqs.~\ref{eq:PostParamsCov} and \ref{eq:PostParamsMean}. By marginalising again,
this time with respect to the model, we derive the posterior distribution of the
unseen data,
\begin{equation}
  \label{eq:MarginaliseModelU}
  \pi^y_{Y}(y') \propto 
  \exp \left[ 
   \frac12 \left(y' - \tilde{y}'\right)^T
   (\tilde{C}'_{Y})^{-1} 
   \left(y' - \tilde{y}'\right)
   \right]\, ,
\end{equation}
where
\begin{align}
  \label{eq:lineone}
  \tilde{C}'_{Y} 
  &= \mathcal{G}' \tilde{C}'_{X} \mathcal{G}'^{T} \\
  \label{eq:linetwo}
  \tilde{y}'
  &= \mathcal{G}' \tilde{u}'\, ,
\end{align}
and we have introduced the variables $\tilde{u}'$ and $\tilde{C}'_{X}$,  
\begin{align}
  \label{eq:ModelPostSequential}
  \tilde{C}_{X}'^{-1} 
  &= \mathcal{G}'^T C_{Y}'^{-1} \mathcal{G}' + \tilde{C}_{X}^{-1} \\
  \label{eq:ModelPostSequentialTwo}
  \tilde{u}' 
  &= \tilde{C}'_{X} \left(
    \mathcal{G}'^T C_{Y}'^{-1} y_0' + \tilde{C}_{X}^{-1} \tilde{u} 
    \right) \, .
\end{align}
The variables $\tilde{C}_{X}$ and $\tilde{u}$ have been defined above. We repeat
their definition here in order to have all the necessary equations collected
together: 
\begin{align}
  \label{eq:linethree}
  \tilde{C}_{X}^{-1}
  &= \mathcal{G}^T C_{Y}^{-1} \mathcal{G} + C_{X}^{-1} \\
  \label{eq:linefour}
  \tilde{u}
  &= \tilde{C}_{X} \left(
    \mathcal{G}^T C_{Y}^{-1} y_0 + C_{X}^{-1} u_0
  \right)\, .
\end{align}
Eqs.~\ref{eq:lineone}, \ref{eq:linetwo}, \ref{eq:ModelPostSequential},
\ref{eq:ModelPostSequentialTwo}, \ref{eq:linethree} and \ref{eq:linefour} yield
the central value and the variance of the posterior distribution of unseen data
as a function of $y_0$, $y_0'$, $C_Y$, $C_Y'$, $u_0$ and $C_X$. 

\paragraph{A comment on non-linear models}

The linear models that we have discussed so far may look over-simplified at
first sight. In practice, it turns out that non-linear models can often be
linearised around the central value of the prior distribution, 
\begin{equation}
  \label{eq:LinU0}
  \fwdobsop(\modelvec) = \fwdobsop(\modelpriorcent) + G \left(\modelvec - \modelpriorcent\right) + \ldots\, ,
\end{equation}
where 
\begin{equation}
  \label{eq:FirstDerU0}
  G^i_\alpha = \left. \frac{\partial \fwdobsop^i}{\partial u_\alpha} \right|_{\modelpriorcent}\, ,
\end{equation}
and we have neglected higher-order terms in the expansion of
$\fwdobsop(\modelvec)$.

If these terms are not negligible, another option is to find the MAP estimator,
and then expand the the forward map around it, which yields equations very
similar to the previous ones, with $\modelpriorcent$ replaced by $u_*$. If the
posterior distribution of $u$ is sufficiently peaked around the
MAP estimator, then the linear approximation can be sufficiently accurate.

\subsection{The infinite-dimensional case}
\label{sec:infin-dimens-case}

In the finite-dimensional case, where the probability measures are specified by
their densities with respect to the Lebesgue measure,
Eq.~(\ref{eq:BayesThmInversePosterior}) can be rephrased by saying  that $\rho$
is the Radon-Nikodym derivative of the probability measure $\mu^\fwdobsop$ with
respect to $\mu_0$, \viz
\begin{align}
  \label{eq:RadonNikodym}
  \frac{d\mu^\fwdobsop}{d\mu^0} (\modelvec) \propto \rho(\obs-\fwdobsop(\modelvec))\, .
\end{align}
Using the fact that the density $\rho$ is a positive function, we can rewrite 
\begin{align}
  \label{eq:PotentialDef}
  \rho(\obs-\fwdobsop(\modelvec)) = \exp\left(-\Phi(\modelvec;\obs)\right)\, ,
\end{align}
and therefore
\begin{align}
  \label{eq:RadonNikodymTwo}
  \frac{d\mu^\fwdobsop}{d\mu^0} (\modelvec) \propto \exp\left(-\Phi(\modelvec;\obs)\right)\, .
\end{align}
In finite-dimensional spaces, the three equations above are just definitions
that do not add anything to the above discussion in terms of probability
densities. Their interest resides in the fact that the last expression,
Eq.~(\ref{eq:RadonNikodymTwo}), can be properly defined when $\modelspace$ is
infinite-dimensional, allowing a rigorous extension of the Bayesian formulation
of inverse problems to the case of infinite-dimensional spaces. 

Summarising the details of probability measure in infinite-dimensional spaces,
is beyond the scope of this work. Adopting instead a heuristic approach, we can
say that a function $f$ is a random function if $f(x)$ is a random variable for
all values of $x$. Since the values of the function at different values of $x$
can be correlated, a random function is fully characterised by specifying the
joint probability densities
\begin{equation}
  \label{eq:RandomFuncJointProb}
  \pi\left(
    f_1, \ldots, f_n; x_1, \ldots x_n
  \right)\, ,
\end{equation}
where $f_i=f(x_i)$, for all values of $n$, and all values of $x_1, \ldots, x_n$.
This infinite set of finite-dimensional densities allows the definition of a
probability measure. 

For a Gaussian random function, these densities only depend on a mean value
function $m(x)$ and a covariance $C(x,x')$. The probability densities for the
variables $f_i$, for any value of $n$ is 
\begin{equation}
  \label{eq:GaussianFunctC}
  \pi\left(f_1, \ldots, f_n; x_1, \ldots, x_n\right)
  \propto \exp \left[ 
      -\frac12 \sum_{ij} \left(f_i - m_i\right) C^{-1}(x_i,x_j) \left(f_j - m_j\right)
    \right]\, .
\end{equation} 
The covariance $C$ is such that
\begin{equation}
  \label{eq:CovFunctInt}
  C(x,x') = \int df\, df'\, \left(f - m(x)\right) \left(f'-m(x')\right)
    \pi\left(f,f';x,x'\right)\,,
\end{equation}
which shows that the two-point probability density determines all the other
distributions. 

\paragraph{Functional linear problems} This formalism allows us to formulate a
Bayesian solution of the inverse problem 
\begin{equation}
  \label{eq:BayesLinearInverse}
  y^i = \int dx\, G^i(x) u(x)\, ,
\end{equation}
where $y^i$ is a discrete set of observables and $u(x)$ is a random function,
with a Gaussian prior with mean $u_0(x)$ and covariance $C_{X}(x,x')$. The
vector of observed values is denoted $y_0$, and we assume that the prior
distribution of $y$ is a Gaussian centred at $y_0$ with covariance $C_{Y}$.

Similarly to the finite-dimensional case, the Bayesian solution yields a
Gaussian random function for the posterior distribution of the solution $u(x)$.
In order to characterise the posterior Gaussian distribution we need explicit
expressions for its mean and its covariance. Introducing the matrix
\begin{equation}
  \label{eq:Smatrix}
  S^{ij} =
  \int dx dx'\, G^i(x) C_{X}(x,x') G^j(x') + C_{Y}^{ij}\, ,
\end{equation}
and its inverse $T^{ij}=\left(S^{-1}\right)^{ij}$, the posterior Gaussian field
is centred at
\begin{align}
  \label{eq:PostMeanFunc}
  \tilde{u}(x) = u_0(x) + 
  \int dx'\, C_{X}(x,x') G^i(x') T^{ij} \left(
    y_0^j - \int dx''\, G^j(x'') u_0(x'') 
  \right)\, ,
\end{align}
which is the expected generalization of the finite-dimensional example discussed
above. Interestingly, this can be rewritten as
\begin{equation}
  \label{eq:TowardsBackus}
  \tilde{u}(x) = u_0(x) + 
  \int dx'\, C_{X}(x,x') \psi(x')\, ,
\end{equation}
where 
\begin{eqnarray}
  \label{eq:PsiDef}
  \psi(x) = G^i(x) \delta y^i\, ,
\end{eqnarray}
and the weighted residuals are given by
\begin{equation}
  \label{eq:DeltaYDef}
  \delta y^i = T^{ij} \left(
  y_0^j - \int dx'\, G^j(x') u_0(x')
  \right)\, .
\end{equation}
Defining 
\begin{equation}
  \label{eq:CapitalPsi}
  \Psi^i(x) = \int dx'\, C_{X}(x,x') G^i(x')\, ,
\end{equation}
the posterior covariance can be written as
\begin{equation}
  \label{eq:PostCovFunc}
  \tilde{C}_{X}(x,x') = 
  C_{X}(x,x') - \Psi^i(x) T^{ij} \Psi^j(x')\, .
\end{equation}

It is instructive to compare the Bayesian result summarised above with the
method proposed by Backus and Gilbert~\cite{BackusGilbert1968} to solve the same
inverse problem. Assuming that there exists an unknown 'true' model $\utrue$,
such that the observed data are
\begin{equation}
  \label{eq:BackStart}
  y_0^i = \int dx\, G^i(x) \utrue(x)\, ,
\end{equation}
we look for an estimate $\uest$ of the true solution in the form
\begin{equation}
  \label{eq:BackAnsatz}
  \uest(x) = Q^i(x) y^i_0\, ,
\end{equation}
so that the problem is now recast as finding the functions $Q^i(x)$.
Using Eq.~\ref{eq:BackStart} we obtain
\begin{equation}
  \label{eq:BackFilter}
  \uest(x) = \int dx' R(x,x') \utrue(x')\, , 
\end{equation}
which states that with a finite amount of data we can only resolve a filtered
version of the true solution. The kernel $R$ is given by
\begin{equation}
  \label{eq:BackKernel}
  R(x,x') = Q^i(x) G^i(x')\, .
\end{equation}
The coefficient functions $Q^i(x)$ can be chosen so that the kernel is as close as possible
to a delta function,
\begin{equation}
  \label{eq:BackDelta}
  R(x,x') \simeq \delta(x,x') ~~ \Longrightarrow ~~
  \uest \simeq \utrue\, .
\end{equation}
Approximating the delta function can be achieved by minimising 
\begin{equation}
  \label{eq:BackDeltaness}
  \int dx'\, \left(
    R(x,x') - \delta(x-x')
  \right)^2\, ,
\end{equation}
which yields
\begin{equation}
  \label{eq:BackSolution}
  Q^i(x) = \left(S^{-1}\right)^{ij} G^j(x)\, ,
\end{equation}
where 
\begin{equation}
  \label{eq:BackSMatrix}
  S^{ij} = \int dx\, G^i(x) G^j(x)\, .
\end{equation}
The interesting observation is that the central value of the Bayesian solution
presented above reduces to the Backus-Gilbert $\uest$ in the case where $u_0$ 
is just white noise and therefore
\begin{equation}
  \label{eq:BackComparison}
  C_{X}(x,x') = \delta(x-x')\, .
\end{equation}
The connections between the Bayesian treatment and the Backus-Gilbert solution
and its regulated variations, deserves further investigations, which we defer to
future studies. Note that the Bayesian solution allows a variety of priors to be
explicitly declared and compared to the Backus-Gilbert solution. 

%% file: closure_test.tex
\section{NNPDF Monte Carlo approach to inverse problems}
\label{sec:closure-test}

In this section we discuss the NNPDF approach to inverse problems, and make
contact explicitly with the formalism laid out in
Sec.~\ref{sec:inverse-problems}. In the Bayesian formulation,
Eq.~\ref{eq:PosteriorModel} gives a quantitative description of how the
information contained in the experimental data propagates into our knowledge of
the space of models. In practice, it should be possible to sample directly from
the posterior distribution, or to find a solution using some kind of
Backus-Gilbert methodology~\cite{BackusGilbert1968}. These new types of approach
are not straightforward and we defer their investigation to further, dedicated
studies. Here we focus instead on the standard NNPDF fitting procedure and
investigate its relation with the Bayesian result. The NNPDF approach generates
an ensemble of fit results, which are supposed to describe the posterior
probability distribution for the model (\ie\ in the space of PDFs) given the
experimental data. In the case of a linear map, we show here that this is
exactly the case: the NNPDF replicas are distributed exactly accroding to the
posterior density that was obtained in the previous section. 

\subsection{Fitting replicas}
\label{sec:fit-reps}

\begin{figure}
    \centering
    \includegraphics[scale=0.8]{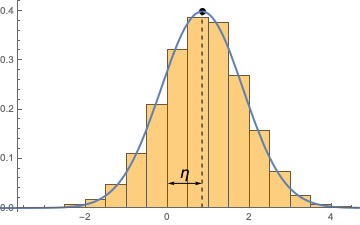}
    \caption{Histogram showing the distribution of $10^4$ replicas generated
    around an experimental value $y_0$ with unit variance. The central value
    $y_0$, which is represented by the solid dot at the centre of the replica
    distribution, is drawn from a Gaussian distribution with unit variance
    centred at the true value $f$, which is assumed to be the origin in this
    plot. The central value $y_0$ is kept fixed for the whole set of
    replicas.\label{fig:DistRep}}
\end{figure}

The approach for generating a sample in model space utilised by NNPDF can
broadly be described as fitting model replicas to pseudo-data replicas. As
discussed in Eq.~\ref{eq:NoisyInverseProblem} the experimental values are
subject to observational noise. If we assume this observational noise to be
multigaussian then the experimental central values, $\obspriorcent$, are given
explicitly by
\begin{equation}
    \label{eq:levelonedata}
    \obspriorcent = \law + \obsnoise,
\end{equation}
where $\law$ is the vector of {\em true} observable values, and the
observational noise is drawn from a Gaussian centred on zero such as in
Eq.~\ref{eq:RhoGauss}, \ie\ $\obsnoise \sim \mathcal{N}(0, \obspriorcov)$ where
$\obspriorcov$ is the experimental covariance matrix. In
Eq.~\ref{eq:levelonedata}, each basis vector corresponds to a separate data
point, and the vector of shifts $\obsnoise$ permits correlations between data
points according to the covariance matrix provided by the experiments. Given the
data, the NNPDF approach is to compute a MAP estimator along the lines discussed
in the previous section, \ie\ finding the model that minimises the $\chi^2$ to
the data. The key difference between the NNPDF approach and the classical MAP
estimator is that instead of fitting the observational data given by
Eq.~\ref{eq:levelonedata}, an ensemble of model replicas are fitted each to an
independently sampled instance of pseudo-data, which is generated by augmenting
$\obspriorcent$ with some noise, $\noise^{\repind}$,
\begin{equation}
    \label{eq:leveltwodata1}
    \pseudodat^{\repind} = \obspriorcent + \noise^{\repind}
    = \law + \obsnoise + \noise^{\repind},
\end{equation}
where $k$ is the replica index and each instance of the noise, $\noise$, is
drawn independently from the same Gaussian from which the observational noise is
drawn from, \ie\ $\noise \sim \mathcal{N}(0, \obspriorcov)$. For each replica
$k$, $\pseudodat^{\repind}$ is a vector in $\real^{\ndata}$. The distribution of
pseudo-data in a simple one-dimensional example is shown in
Fig.~\ref{fig:DistRep}. Note that, if we were to repeat this construction
multiple times, the true value $f$ would be within a 1$\sigma$ interval centred
at $y_0$ with a 68\% probability.

The parameters for each model replica maximise the likelihood evaluated on the
corresponding pseudo-data. We can think of this approach as a special case of
MAP estimation, as described in Eq.~\ref{eq:MAP}, where there is no model
prior that regulates the likelihood. Another way of viewing this is to take
$\modelpriorcov^{-1} \to 0$ in Eq.~\ref{eq:MAP}, as was done to obtain the
result in Eq.~\ref{eq:NoPriorLinModel}. Either way, there is no prior
information about the model. The parameterisation of the model is fixed, so the
model space is the space of parameters $\modelvec \in \real^{\nmodel}$. In
$\real^{\nmodel}$, we find the parameters which minimise the $\chi^2$ between
the predictions from the model and the corresponding pseudo-data
$\pseudodat^{\repind}$
\begin{equation}\label{eq:NNPDFLikelihood}
    \begin{split}
        \modelvecrep &= \arg\min_{\modelvec^{\repind}} \repchis \\
        &= \arg\min_{\modelvec^{\repind}} \sum_{ij}
        \left( \fwdobsop(\modelvec^{\repind}) - \pseudodat^{\repind} \right)^T
        \obspriorcov^{-1}
        \left( \fwdobsop(\modelvec^{\repind}) - \pseudodat^{\repind} \right) \, ,
    \end{split}
\end{equation}
where, as usual, minimising the $\chi^2$ is equivalent to maximising the
likelihood, $\likelihood$, since $\chi^2 \equiv -\log{\likelihood}$.

As a final note: since we do not include the model prior, overall normalisations
can be omitted in Eq.~\ref{eq:NNPDFLikelihood}. It is clear however that if we
were including a model prior in our MAP, it is important that the relative
normalisation between the likelihood function and the model prior is clearly
specified.

\subsection{Fluctuations of fitted values}
\label{sec:fluct-fit-values}

It is not immediately obvious that the MC methodology, maximising the likelihood
on an ensemble of pseudo-data replicas, guarantees that the model replicas are
indeed sampled from the posterior distribution of parameters given data as
described \eg\ in Eq.~\ref{eq:PosteriorModel}. In order to investigate this
issue, we will again consider a model whose predictions are linear in the model
parameters, so that the posterior distribution of model parameters can be
computed explicitly. DIS observables can be seen as an example of a linear
model. Let us assume that the PDF are parametrized by their values at selected
values of $x_i$, so that $u$ is a finite-dimensional vector 
\begin{equation}
    \label{eq:Uvector}
    u_i = u(x_i)\, , \quad i = 1, \ldots, \nmodel\, .    
\end{equation}
The observables are then computed by taking a matrix-vector multiplication of
the vector $u$ by a $\ndata \times \nmodel$ matrix, $F_{ij}$, which is called FK
table in the NNPDF jargon, 
\begin{equation}
    \label{eq:FKTabLinVec}
    y_i = \sum_{j=1}^{\nmodel} F_{ij} u_j\, ,
\end{equation}
where $i=1,\ldots,\ndata$. In this case the forward map coincides exactly with
the FK table, $\linmap_{ij} = F_{ij}$ and the sum in Eq.~\ref{eq:FKTabLinVec}
approximates the convolution in Eq.~\ref{eq:DISExample}. However it should be
clear that the discussion below applies to any model whose forward map can be
approximated as Eq.~\ref{eq:FKTabLinVec}, like a linear approximation of
neural networks~\cite{ADVANI2020428}. Let us consider for instance the actual
NNPDF parametrization, where the value of the PDF is given by a neural network
whose state is determined by a set of weights $\theta$. In this case $X$ is the
space of weights, the model will be specified by a vector $\theta$ in the space
of weights, and we are going to continue using $u$ to denote the PDF, so that 
\begin{equation}
    \label{eq:WeightsParam}
    u_i = u(x_i; \theta)\, .
\end{equation}
If the parameters have a sufficiently narrow distribution around some value
$\bar\theta$, then we can expand the expression for the observables:
\begin{align}
    y_i  &= 
        \sum_j F_{ij} \left[u(x_j; \bar\theta)
         + \sum_k \left.\frac{\partial u}{\partial \theta_k}\right|_{x_j,\bar\theta} 
         \left(\theta - \bar\theta\right)_k
        \right]
\end{align}
and therefore
\begin{align}
    y_i - \bar{y}_i &=
         \sum_k \linmap_{ik} \Delta\theta_k
        \,,         
\end{align}
where 
\begin{align}
    \linmap_{ik} &= 
        \sum_j F_{ij} \left.\frac{\partial u}{\partial \theta_k}\right|_{x_j,\bar\theta}\, , \\
    \bar{y}_i &= \sum_j F_{ij} u(x_j; \bar\theta)\, , \\
    \Delta\theta_k &= \theta_k - \bar\theta_k\, .
\end{align}

In order to get an exact analytical solution for the linear model, we
additionally require $\linmap$ to have linearly independent rows, and therefore
$\linmap \obspriorcov \linmap^T$ is invertible. As discussed in
Sec.~\ref{sec:inverse-problems}, in the absence of prior information on the
model, the posterior distribution of model parameters is a Gaussian with mean
and covariance given by Eqs.~\ref{eq:NoPriorLinModel} and
\ref{eq:NoPriorLinModelCov}.

Let us now stick to the parametrization in Eq.~\ref{eq:Uvector} and let us
deploy the NNPDF Monte Carlo method to fitting model replicas, then in the case
under study $\arg\min_{\modelvec^{\repind}} \repchis$ is found analytically by
imposing that the derivative of $\repchis$ with respect to the model parameters
is zero, i.e.
\begin{equation}
    \begin{split}
        \label{eq:MAPEstLinModel}
        \modelvecrep &= (\linmap^T \obspriorcov^{-1} \linmap)^{-1}
        \left(
            \linmap^T \obspriorcov^{-1} \obspriorcent +
            \linmap^T \obspriorcov^{-1} \noise^{\repind}
        \right) \, .
    \end{split}
\end{equation}
Eq.~\ref{eq:MAPEstLinModel} shows that $\modelvec_*$ is a linear combination of
the Gaussian variables $\noise$, and therefore is also a Gaussian variable. Its
probability density is then completely specified by the mean and covariance,
which can be calculated explicitly, given that the probability density for
$\noise$ is known:
\begin{align}
    \emodel{\modelvec_*} &=
    \modelpostcent = (\linmap^T \obspriorcov^{-1} \linmap)^{-1} \linmap^T
    \obspriorcov^{-1} \obspriorcent\, , \\
    {\rm cov}(\modelvec_*) &= \modelpostcov = (\linmap^T \obspriorcov^{-1} \linmap)^{-1} \, .
\end{align}
These two equations show that, under the assumptions specified above,
$\modelvec_* \sim \mathcal{N}(\modelpostcent, \, \modelpostcov)$. In other
words, when the model predictions are linear in the model parameters, the NNPDF
MC method is shown to produce a sample of models exactly distributed according
to the expected posterior distribution of model parameters given the data. When
we fit PDFs, parameterised as deep fully connected neural networks, to data
which includes hadronic observables, it is clear that the forward map is
non-linear, and therefore this proof does not strictly apply. As discussed
above, even for non-linear models we can make a linear approximation of the
forward map provided that we are expanding around the MAP estimator. This means
the NNPDF MC methodology should reproduce the posterior distribution of the
model given the data, at least close to $\modelpostcent$, the central value of
the fitted replicas. Furthermore, by fluctuating the data and fitting the
replicas, the fluctuations in data space are propagated to model space
non-linearly. So even for non-linear problems, the NNPDF MC methodology can be
expected to produce a sample of models which are at least approximately
distributed according to the posterior model distribution. It remains to be
shown, however, that further away from the MAP estimator the approximation holds
despite the non-linear dependence of the model replicas on the data
uncertainities.

\subsection{Closure test}
\label{sec:closure-test-intro}

The concept of the closure test, which was first introduced in
Ref.~\cite{nnpdf30}, is to construct artificial data by using a known
pre-existing function to generate the {\em true} observable values, $\law$. One
way of achieving this is by choosing a model $\lawmodel$ and then compute $\law
= \fwdobsop(\lawmodel)$. Then the experimental central values are artificially
generated according to Eq.~\ref{eq:levelonedata}, where the observational noise
is pseudo-randomly generated from the assumed distribution. In
Ref.~\cite{nnpdf30}, $\law$ is referred to as level 0 (L0) data and
$\obspriorcent$ is referred to as level 1 (L1) data. Finally, if we use the
NNPDF MC method to fit artificially generated closure data, the pseudo-data
replicas that are fitted by the model replicas are referred to as level 2 (L2)
data.

The outcome of the fit is then compared with the known {\em true} value to check
the precision and accuracy of the fitting procedure. Finding quantitative
estimators that allow us to characterise the quality of the closure fit is one
of the main problems that need to be addressed. We will discuss a new class of
estimators in the next section. 

Note that in a closure test, the assumed prior of the data is fully consistent
with the particular instance of observed central values, $\obspriorcent$: by
construction, there are no inconsistent data sets. In the original closure test
of Ref.~\cite{nnpdf30} there was also no modelisation uncertainty, the true
observable values were assumed to be obtained by applying the forward map
$\fwdobsop$ to a vector in model space $\lawmodel$. It is worth noting that the
assumption of zero modelisation uncertainties is quite strong and likely
unjustified in many areas of physics. In the context of fitting parton
distribution functions there are potentially missing higher order uncertainties
(MHOUs) from using fixed order perturbative calculations as part of the forward
map. MHOUs have been included in parton distribution fits \cite{NNPDF:2019vjt,
AbdulKhalek:2019ihb} and in the future these should be included in the closure
test, however this is beyond the scope of the study presented here, since MHOUs
are still not included in the NNPDF methodology by default. In the results
presented in the rest of this paper we do include nuclear and deuteron
uncertainties, as presented in \cite{Ball:2018twp, Ball:2020xqw, AbdulKhalek:2020yuc}, since they are
to be included in NNPDF fits by default. Extensive details for including
theoretical uncertainties, modelled as theoretical covariance matrices can be
found in those references. For the purpose of this study the modelisation
uncertainty is absorbed into the prior of the data, since
\begin{equation}
    \obspriorcent = \fwdobsop(\modelvec) + \obsnoise + \delta
\end{equation}
where $\delta \sim \mathcal{N}(0, \cov^{\rm theory})$. As long as the
modelisation uncertainty is independent of the data uncertainty, we can absorb
$\delta$ into $\obsnoise$ by modifying the data prior: $\obsnoise \sim
\mathcal{N}(0, C + C^{\rm theory})$. In doing that, we must also update the
likelihood of the data given the model to use the total covariance $(C + C^{\rm
theory})$. From now onwards we will omit $C^{\rm theory}$ because it is implicit
that we always sample and fit data using the total covariance matrix which
includes any modelisation uncertainty we currently take into account as part of
our methodology.

Mapping the closure test procedure to the quantities used in the Bayesian
treatment presented in the previous section will allow us to derive a number of
analytical results in Sec.~\ref{Sec:LinearMapEstimators}.

%% file: estimators.tex
\section{Data space estimators} 
\label{sec:ClosureEstimators}

In order to perform a quantitative analysis of the results obtained in the
closure tests, we discuss several estimators, which are computed from the
outcome of the closure test fits. These results depend on the pseudo-data that
have been generated and therefore are stochastic variables which can fluctuate.
The values of the estimators on a single replica will not tell anything about
the quality of our fits: we need to understand their probability distributions
in order to validate our fitting procedure. We begin this section by defining
estimators in data space, \ie\ estimators that are computed from the model
predictions for a set of experimental data points. Having defined the
estimators, we define criteria to characterise faithful uncertainties. We
conclude this section with a discussion of the predictions that can be obtained
for these estimators in the case of a linear model, where analytical
calculations can be performed. As we already saw in the previous section, the
analytical results cannot be applied directly to the NNPDF fits, but they are
useful examples that illustrate the expected behaviour of these quantities.

\subsection{Deriving the data space estimators}
\label{sec:ClosureEstimatorsDerivation}

For a given model $\modelvecrep$, obtained from fitting the $k$-th replica, we
start by defining the model error as the $\chi^2$ between the model predictions
and some data central values $\testset{\obspriorcent}$, normalised by the number
of data points
\begin{equation}
    \label{eq:chi2kereponerep}
    \frac{1}{\ndata} 
        \left( \testset{\fwdobsop}\left(\modelvecrep\right) - \testset{\obspriorcent} \right)^T
        \testset{\obspriorcov}^{-1}
        \left( \testset{\fwdobsop}\left(\modelvecrep\right) - \testset{\obspriorcent} \right)\, ,
\end{equation}
and we purposely denoted the data which the model error is evaluated on as
$\testset{\obspriorcent}$, as opposed to the training data $\obspriorcent$,
which is used to determine the model parameters. The corresponding covariance is
denoted $\testset{\obspriorcov}$. Note that in Eq.~\ref{eq:chi2kereponerep},
$\testset{\obspriorcent}$ is a stochastic variable, but also $\modelvecrep$ is a
stochastic variable, with its pattern of fluctuations, since the fitted model
depends on the data $\pseudodat^{\repind}$ that enter the fit. We define the
model error $\eout$ on the set of data $\testset{\obspriorcent}$ by taking the
average over the models,
\begin{equation}
    \label{eq:chi2kerep}
    \eout = \frac{1}{\ndata} \emodel{
        \left( \testset{\fwdobsop}\left(\modelvecrep\right) - \testset{\obspriorcent} \right)^T
        \testset{\obspriorcov}^{-1}
        \left( \testset{\fwdobsop}\left(\modelvecrep\right) - \testset{\obspriorcent} \right)
    }\, ,
\end{equation}
where we defined the expectation value over the ensemble of model replicas as
\begin{equation}
    \emodel{x} \equiv \frac{1}{\nreps} \sum_{k=1}^{\nreps} x^{(k)} \, .
\end{equation}
We could of course set $\testset{\obspriorcent} = \obspriorcent$ and evaluate
the model performance on the fitted data however, as is common in machine
learning literature, we intend to use a separate set of test data. Ideally we
would choose $\testset{\obspriorcent}$ such that $\testset{\obspriorcent}$ and
$\obspriorcent$ are statistically independent, as in
Eq.~\ref{eq:JointIndepDataPrior}. This is achieved by choosing the split such
that the experimental covariance matrix is block diagonal:
\begin{equation}
    \modelpriorcov^{\rm total} =
    \begin{bmatrix}
        \modelpriorcov  & 0  \\ 
        0  & \testset{\modelpriorcov}  \\ 
    \end{bmatrix}\, .
\end{equation}

It is useful to perform a decomposition of Eq.~\ref{eq:chi2kerep}, following
usual manipulations of the likelihood function associated with least-squares
regression in~\cite{mlforphysics}. Least-squares regression is a special case of
minimum likelihood estimation, where the uncertainty on each data point is equal
in magnitude and uncorrelated. Here we review the decomposition in the more
general framework of data whose uncertainty is multigaussian. Starting with
Eq.~\ref{eq:chi2kerep} (evaluated on the ideal test data), we can complete the
square
\begin{equation}
    \begin{split}
    \label{eq:EoutDecomposition}
        &\eout = \frac{1}{\ndata} \biggl\{ \emodel{
            \left( \testset{\fwdobsop}\left(\modelvecrep\right) - \testset{\law} \right)^T
            \testset{\obspriorcov}^{-1}
            \left( \testset{\fwdobsop}\left(\modelvecrep\right) - \testset{\law} \right)
        } + \\
        &+ \emodel{
            \left( \testset{\law} - \testset{\obspriorcent} \right)^T
            \testset{\obspriorcov}^{-1}
            \left( \testset{\law} - \testset{\obspriorcent} \right)
        }+ \\
        &+ 2 \emodel{
            \left( \testset{\fwdobsop}\left(\modelvecrep\right) - \testset{\law} \right)^T
            \testset{\obspriorcov}^{-1}
            \left(\testset{\law} - \testset{\obspriorcent} \right)
        }\biggr\}\, .
    \end{split}
\end{equation}
Let us now discuss these terms one by one, starting from the last two. The
second term is the shift associated with evaluating the model error on noisey
test data and the final term is a cross term which we will deal with later. We
therefore focus next on further decomposing the first term,
\begin{equation}
    \begin{split}
        &\emodel{
            \left( \testset{\fwdobsop}\left(\modelvecrep\right) - \testset{\law} \right)^T
            \testset{\obspriorcov}^{-1}
            \left( \testset{\fwdobsop}\left(\modelvecrep\right) - \testset{\law} \right)
        } = \\
        &= \emodel{
            \left( \testset{\fwdobsop}\left(\modelvecrep\right) - 
            \emodel{\testset{\fwdobsop}\left(\modelvecrep\right)} \right)^T
            \testset{\obspriorcov}^{-1}
            \left( \testset{\fwdobsop}\left(\modelvecrep\right) - 
            \emodel{\testset{\fwdobsop}\left(\modelvecrep\right)} \right)
        } + \\
        &+ \left( \emodel{\testset{\fwdobsop}\left(\modelvecrep\right)} - \testset{\law} \right)^T
        \testset{\obspriorcov}^{-1}
        \left( \emodel{\testset{\fwdobsop}\left(\modelvecrep\right)} - \testset{\law} \right)\, ,
    \end{split}
\end{equation}
where we have used the fact that the second term is constant across replicas and
the cross term that arises in this decomposition is zero when the expectation
value across replicas is taken. The first term in this expression we call the
{\em variance} and the second term is the {\em bias}.

As previously mentioned $\eout$ should be considered a stochastic estimator, in
theory we could take the expectation value across training data $\obspriorcent$
and test data $\testset{\obspriorcent}$, the latter of which cancels the cross
term in Eq.~\ref{eq:EoutDecomposition}. The final result of that would be
\begin{equation}\label{eq:ExpectedBiasVariance}
    \mathbf{E}_{\obspriorcent, \testset{\obspriorcent}}[\eout] =
    \mathbf{E}_{\obspriorcent}[{\rm bias}] + 
    \mathbf{E}_{\obspriorcent}[{\rm variance}] +
    \mathbf{E}_{\testset{\obspriorcent}}[{\rm noise}]\, .
\end{equation}
We are not interested in the observational noise term, since it is independent
of the model and in the limit of infinite test data
$\mathbf{E}_{\testset{\obspriorcent}}[{\rm noise}] \to 1$. The two estimators of
interest are independent of the test data, and therefore we only need to take
the expectation value over the training data. In practical terms, this can be
achieved by running multiple closure fits, each with a different observational
noise vector $\obsnoise$, and taking the average i.e.
\begin{equation}
    \label{eq:average_over_training_data}
    \mathbf{E}_{\obspriorcent}[ x ] = \frac{1}{\nfits} \sum_{j=1}^{\nfits} x.
\end{equation}
Clearly this is resource intensive, and requires us to perform many fits. In
NNPDF3.0 \cite{nnpdf30}, single replica proxy fits were used to perform a study
of the uncertainties. Here we have expanded the data-space estimators used in
the closure fits and also will be using multiple full replica fits to calculate
various expectation values - made possible by our next generation fitting code.

\subsection{Geometric Interpretation}

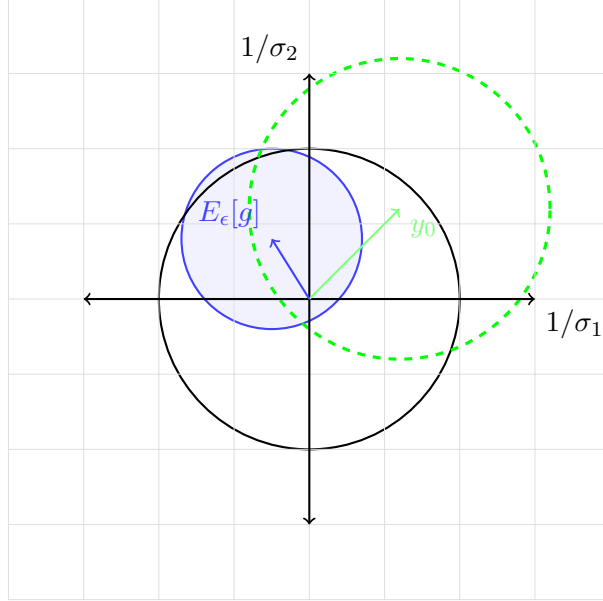
\begin{figure}[h!]
    \centering
    \begin{tikzpicture}
        \draw[blue!75, fill=blue!5, thick] (-0.5, 0.8) circle (1.2 cm);
        \draw[thick] (0,0) circle (2 cm);
        \draw[green, very thick, dashed] (1.2,1.2) circle (2 cm);
        \draw[step=1cm,gray!25!,very thin] (-4,-4) grid (4,4);
        \draw[thick,<->] (-3,0) -- (3,0) node[anchor=north west] {$1/\sigma_1$};
        \draw[thick,<->] (0,-3) -- (0,3) node[anchor=south east] {$1/\sigma_2$};
        \draw[green!50,thick,->] (0,0) -- (1.2,1.2) node[anchor=north west] {$\obspriorcent$};
        \draw[blue!75,thick,->] (0,0) -- (-0.5, 0.8) node[anchor=south east] {$E_\epsilon[g]$};
        \end{tikzpicture}
    \caption{Example of geometric interpretation of closure test estimators. The
    origin is the true observable values for each data point. The level one data
    (or experimental central values) are shifted away from this by $\obsnoise$.
    In this example the covariance matrix is diagonal, so the eigenvectors
    correspond to the two data points, the square root of the eigenvalues are
    simply the standard deviation of those points. This is without loss of
    generality because any multivariate distribution can be rotated into a basis
    which diagonalises the covariance matrix. The 1-sigma observational noise
    confidence interval is a unit circle centered on the origin. Some closure
    estimators can be understood as l2-norms of the vectors connecting points,
    i.e the bias is the l2-norm of the vector from the origin to the central
    value of the predictions.}
    \label{fig:diagram2destimators}
\end{figure}

It is possible to interpret the relevant data space estimators geometrically, by
considering a coordinate system where each basis vector corresponds to an
eigenvector of the experimental covariance matrix normalised by the square root
of the corresponding eigenvalue. An example of this is given in
Fig.~\ref{fig:diagram2destimators}, where for simplicity we have considered a
system with just two data points, \ie\ a two-dimensional data space, with a
diagonal covariance. The origin of the coordinate system is the true value of
the observable. The observational noise in these coordinates corresponds to a
unit circle centred in the origin as shown in
Fig.~\ref{fig:diagram2destimators}. If the experimental covariance is faithful,
there is a 68\%  probability that the experimental value $y_0$ is within this
unit circle. Fig.~\ref{fig:diagram2destimators} shows one possible instance of
$y_0$. Repeating the entire fit procedure multiple times requires generating new
sets of experimental data $y_0$. The average over $y_0$ mentioned above, is
precisely the average over multiple fits, restarting the procedure each time
from a new instance of $y_0$.

For a given $y_0$ the replicas are generated as a set of points Gaussianly
distributed around it and therefore, in the limit of a large number of replicas,
68\% of them will fall within a unit circle centred in $y_0$. This is the dashed
circe in the figure. Clearly there is also a 68\% probability that the true
value (\ie\ the origin in our plot) is inside this second circle. The model
predictions, one for each replica, are then a set of points, whose mean is
$E_\epsilon[g]$. The mean squared radius of those points is what we call the
variance. The bias is the l2-norm of the vector between the origin and the mean
of the model predictions. 

A faithful representation of the errors requires that the true value, \ie\ the
origin of the coordinate system in our figure, has 68\% probability of being
within 1$\sigma$ from the central value of the fit, which is given by
$E_\epsilon[g]$. Looking at the figure again, the probability for the origin to
be inside the shaded circle must be 68\%. We will discuss faithful errors in
more detail in the next subsection.

\subsection{Faithful uncertainties in data space}

The two closure estimators of interest, bias and variance, can be used to
understand faithful uncertainties in a practical sense. If we return to
Eq.~\ref{eq:ExpectedBiasVariance} we can examine both estimators in a more
detail.

\paragraph{Variance}

The {\em variance} in the above decomposition refers to the variance of the
model predictions in units of the covariance
\begin{equation}
    \label{eq:VarDef}
    \begin{split}
        \var &= \frac{1}{\ndata}\mathbf{E}_{\{ \modelvec_* \}} \Big[ \\
            &\left( \testset{\fwdobsop}\left(\modelvecrep\right) - 
            \emodel{\testset{\fwdobsop}\left(\modelvecrep\right)} \right)^T
            \testset{\obspriorcov}^{-1}
            \left( \testset{\fwdobsop}\left(\modelvecrep\right) - 
            \emodel{\testset{\fwdobsop}\left(\modelvecrep\right)} \right)
        \Big] \, ,
    \end{split}
\end{equation}
which can be interpreted as the model uncertainty in the space of the test data.
It is instructive to rephrase Eq.~\ref{eq:VarDef} as
\begin{equation}
    \label{eq:VarDefalternative}
    \var = \frac{1}{\ndata} {\rm Tr} \left[ \covrep \testset{\obspriorcov}^{-1} \right],
\end{equation}
where 
\begin{equation}
    \label{eq:CovRep}
    \covrep = 
    \emodel{
            \left( \testset{\fwdobsop}\left(\modelvecrep\right) - 
            \emodel{\testset{\fwdobsop}\left(\modelvecrep\right)} \right)
            \left( \testset{\fwdobsop}\left(\modelvecrep\right) - 
            \emodel{\testset{\fwdobsop}\left(\modelvecrep\right)} \right)^T
        }
\end{equation}
is the covariance matrix of the predictions from the model replicas. Note that
we can rotate to a basis where $\testset{\obspriorcov}$ is diagonal,
\begin{equation}
    \label{eq:InvCovPrimeDiag}
    \left(\testset{\obspriorcov}^{-1} \right)_{ij} = \frac{1}{\left(\testset{\sigma}_i\right)^2} 
    \delta_{ij}\, ,
\end{equation}
then we can rewrite Eq.~\ref{eq:VarDefalternative} as 
\begin{equation}
    \label{eq:VarianceInterpretation}
    \var = \frac{1}{\ndata}\, \sum_i \frac{\covrep_{ii}}{\left(\testset{\sigma}_i\right)^2}\, .
\end{equation}
The numerator in the right-hand side of the equation above is the variance of
the theoretical prediction obtained from the fitted replicas, while the
denominator is the experimental variance, the average is now taken over
eigenvectors of the experimental covariance matrix. Note that this ratio does
not need to be equal to one, the model error on a given point can be smaller
than the experimental one, since all theoretical estimates are correlated by the
underlying theoretical law. 

\paragraph{Bias}

Similarly, the {\em bias}\ is defined as the difference between the expectation
value of the model predictions and the true observable values in units of the
covariance, \ie 
\begin{equation}
    \label{eq:BiasDef}
    \bias = \frac{1}{\ndata}
    \left( \emodel{\testset{\fwdobsop}\left(\modelvecrep\right)} - \testset{\law} \right)^T
    \testset{\obspriorcov}^{-1}
    \left( \emodel{\testset{\fwdobsop}\left(\modelvecrep\right)} - \testset{\law} \right)\, .
\end{equation}
The smaller the bias, the closer the central value of the predictions is to the
underlying law. In Eq.~\ref{eq:ExpectedBiasVariance}, the expectation value is
taken across the prior distribution of the training data, which yields
\begin{equation}
    \mathbf{E}_{\obspriorcent}[{\rm bias}] = \frac{1}{\ndata}
    {\rm Tr} \left[ \covcent \testset{\obspriorcov}^{-1} \right]\, ,
\end{equation}
where we have introduced $\covcent$ as the covariance of the expectation value
of the model predictions,
\begin{equation}
    \label{eq:CovCentDef}
    \covcent = 
    \mathbf{E}_{\obspriorcent}\left[
        \left( \emodel{\testset{\fwdobsop}\left(\modelvecrep\right)} - \testset{\law} \right)
        \left( \emodel{\testset{\fwdobsop}\left(\modelvecrep\right)} - \testset{\law} \right)^T   
    \right]\, .
\end{equation}
The point is that the bias on the test data is a stochastic variable which
depends on the central value of the training data through $\modelvecrep$. The
matrix $\covcent$ describes the fluctuations of the central value of the model
prediction around the true observable values as we scan different realisations
of the training data. 

It is important to stress the difference between variance and bias. In the case
of the variance, we are looking at the fluctuations of the replicas around their
central value for fixed $\obspriorcent$. This is related to the ensemble of
model replicas we provide as the end product of a fit and can be calculated when
we have one instance of $\obspriorcent$, provided by the experiments. In the
case of the bias we consider the fluctuations of the central value over replicas
around the true theoretical prediction as the values of $\obspriorcent$
fluctuate around $\law$. This latter procedure is only possible in a closure
test, where the underlying true observable is known. The bias as defined here
yields an estimate of the fluctuations of the MAP estimator if we could do
multiple independent instances of their measurements from each experiment.

\paragraph{Bias-variance ratio}

Finally, the {\em bias-variance ratio} is defined as
\begin{equation}
    \label{eq:RatioDef}
    \biasvarratio \equiv \sqrt{\frac{
        \mathbf{E}_{\obspriorcent}[ \bias ]}{
            \mathbf{E}_{\obspriorcent}[ \var ]}}\, ,
\end{equation}
where we have taken the square root, since bias and variance are both mean
squared quantities. The value of $\biasvarratio$ yields a measurement of how
much uncertainties are over or under estimated. If the uncertainties are
completely faithful, then $\biasvarratio = 1$. We note that $\biasvarratio$ is
not completely general: it is not a measure defined in model space and depends
on the choice of test data. Therefore it only gives {\em local} information on
the model uncertainties. If the distribution of the expectation value of model
predictions is gaussian centered on the true observable values, with covariance
$\covcent$ and the distribution of the model replicas is also gaussian, with
covariance $\covrep$ then model uncertainties are faithful if
\begin{equation}\label{eq:IdealRatioDef}
    \covcent {\covrep}^{-1} = 1.
\end{equation}
The difficulty with calculating Eq.~\ref{eq:IdealRatioDef} comes from the fact
that $\covrep$ is likely to have large correlations which would lead it to be
singular or ill-conditioned. As a result, any error estimating $\covrep$ from
finite number of replicas could lead to unstable results. $\biasvarratio$
overcomes this instability by taking the ratio of the average across test data
of these matrices, in units of the experimental covariance matrix. There may
still be large relative errors for smaller eigenvalues of $\covrep$, but these
should not lead to instabilities in $\biasvarratio$ unless they correspond to
directions with very low experimental uncertainty. As an extra precaution, we
shall estimate an uncertainty on $\biasvarratio$ by performing a bootstrap
sample on fits and replicas.

\paragraph{Quantile statistics}
\label{sec:QuantileStatistics}

When the closure test was first presented in \cite{nnpdf30}, there was an
estimator introduced in the space of PDFs which also aimed to estimate
faithfulness of PDF uncertainties using the combined assumption of Gaussian PDF
uncertainties and quantile statistics, called $\xi_{1\sigma}$. Here we can
define an analogous expression in the space of data,
\begin{equation}
    \label{eq:XiDataDef}
    \xisigdat{n} = 
        \frac{1}{\ndata} \sum_{i}^{\ndata} 
        \frac{1}{\nfits} \sum_{l}^{\nfits}
            I_{[-n \testset{\sigma}_i^{(l)}, n \testset{\sigma}_i^{(l)}]}
            \left( \emodel{\testset{\fwdobsop}_i}^{(l)} - \testset{\law}_i \right),
\end{equation}
where $\testset{\sigma}_i^{(l)} = \sqrt{\covrep_{ii}}$ is the standard deviation
of the theory predictions estimated from the replicas of fit $l$ and $I_{[a,
b]}(x)$ is the indicator function, which is one when $a \leq x \leq b$ and zero
otherwise. In other words, $\xisigdat{n}$ is counting how often the difference
between the prediction from the MAP estimator and the true observable value is
within the $n\sigma$-confidence interval of the replicas, assuming they're
Gaussian. Since $\covrep$ is primarily driven by the replica fluctuations, we
assume that it is roughly constant across fits, or independent upon the specific
instance of observational noise. This allows us to write $\xisigdat{n}$ for a
specific data point in the limit of infinite fits, each to a different instance
of the data as
\begin{equation}
    \label{eq:XiIExpecVel}
        \xisigdati{n} =
            \int_{-\infty}^{\infty} I_{[-n \testset{\sigma}_i, n \testset{\sigma}_i]}\,
            \left( \emodel{\testset{\fwdobsop}_i}^{(l)} - \testset{\law}_i \right) 
            \rho(\obsnoise) \, 
            {\rm d}(\obsnoise) \, ,
\end{equation}
where $\emodel{\fwdobsop_i}^{(l)}$ has implicit conditional dependence on
$\obsnoise$. If the distribution of \linebreak
$\emodel{\testset{\fwdobsop}_i}^{(l)} - \testset{\law}_i$ is Gaussian, centered
on zero, we can define ${\testset{ \hat{\sigma} }}_i = \sqrt{\covcent_{ii}}$. In
this case
\begin{equation}
    \label{eq:expectedxi}
    \xisigdati{n} =
    \erf \left( \frac{n \testset{\sigma}_i}{\testset{\modelstd}_i \sqrt{2}}\right),
\end{equation}
which is simply the standard result of integrating a gaussian over some finite
symmetric interval.

The analogy between $\biasvarratio$ and $\xisigdat{n}$ is clear, the ratios of
uncertainties are both attempts to quantify Eq.~\ref{eq:IdealRatioDef} whilst
keeping effects due to using finite statistics under control. Whilst with
$\biasvarratio$ we take the average over test data before taking the ratio,
$\xisigdat{n}$ instead takes the ratio of the diagonal elements - ignoring
correlations. Since the predictions from the model will be compared with
experimental central values, taking into account experimental error, we find it
more natural to calculate $\xisigdat{n}$ in the basis which diagonalises the
experimental covariance of the test data as in Eq.~\ref{eq:InvCovPrimeDiag}. If
we assume that in this new basis, that both
$\frac{\covrep_{ii}}{\left(\sigma'_i\right)^2}$ and
$\frac{\covcent_{ii}}{\left(\sigma'_i\right)^2}$ are approximately constant for
all eigenvectors of the experimental covariance matrix, then we recover the
approximation
\begin{equation}\label{eq:CompareXiRatio}
    \xisigdat{n} \sim \erf \left( \frac{ n\biasvarratio}{\sqrt{2}} \right).
\end{equation}
Whilst it is clear that Eq.~\ref{eq:CompareXiRatio} is reliant on a fair few
assumptions which may not hold, we will use the comparison of $\xisigdat{n}$ with
$\biasvarratio$ to consider how valid these assumptions may be.

\subsection{Closure estimators - Linear problems}
\label{Sec:LinearMapEstimators}

Once again we return to the linear model framework set out in
Sec.~\ref{sec:fluct-fit-values}. We can perform an analytical closure test in
this framework, and check our understanding of the closure estimators. Consider
the true observable values for the test data are given by
\begin{equation}
    \label{eq:LinearLawMap}
    \testset{\law} = \testset{\fwdobsop} \lawmodel
\end{equation}
where $\lawmodel \in \modelspace$, and we assume that the number of (non-zero)
parameters in the underlying law is less than or equal to the number of
parameters in the model, $\nlaw \leq \nmodel$. Using the previous results from
Sec.~\ref{sec:fluct-fit-values}, we can write down the difference between the
true observables and the predictions from the MAP estimator (or the expectation
of the model predictions across model replicas - in the linear model these are
the same)
\begin{equation}
    \begin{split}
        \emodel{\testset{\fwdobsop}\left(\modelvecrep\right)} - \testset{\law} &=
        \testset{\linmap} (\modelpostcent - \lawmodel ) \\
        &= \testset{\linmap} \modelpostcov \linmap^T \obspriorcov^{-1} \, \obsnoise \, ,
    \end{split}
\end{equation}
where we recall that $\linmap$ is the forward map to the training observables
and $\obspriorcent$ are the corresponding training central values. Because the
training observables do not necessarily coincide with the data used to compute
the estimators, we have two different maps $\linmap$ and $\linmap'$. Calculating
the covariance across training data of
$\emodel{\testset{\fwdobsop}\left(\modelvecrep\right)} - \testset{\law}$ gives
\begin{equation}
    \covcent = \testset{\linmap} \modelpostcov \testset{\linmap}^T \, ,
\end{equation}
so the full expression for $\mathbf{E}_{\obspriorcent}[{\rm bias}]$ is given by
\begin{equation}\label{eq:BiasLinearModel}
    \mathbf{E}_{\obspriorcent}[{\rm bias}] = \frac{1}{\ndata}
    {\rm Tr} \left[
        \testset{\linmap} \modelpostcov \testset{\linmap}^T
        \testset{\obspriorcov}^{-1}
    \right].
\end{equation}
We note that if the test data is identical the data the model was fitted on, we
recover an intuitive result $\mathbf{E}_{\obspriorcent}[{\rm bias}] =
\frac{\nmodel}{\ndata}$. In the case of a polynomial model, where the parameter
of the model are the coefficients of the polynomial function, the maximum value
which $\nmodel$ can take whilst $\linmap$ still has linearly independent rows is
$\ndata$ and in this case the $\mathbf{E}_{\obspriorcent}[{\rm bias}]$ takes its
maximum value of 1. The central predictions from the model exactly pass through
each data point.

We can perform a similar exercise on the model replica predictions. The
difference between the predictions from model replica $\repind$ and the
expectation value of the model predictions is
\begin{equation}
    \begin{split}
        \testset{\fwdobsop}\left(\modelvecrep\right) -
        \emodel{\testset{\fwdobsop}\left(\modelvecrep\right)} &=
        \testset{\linmap} (\modelvecrep - \modelpostcent) \\
        &= \testset{\linmap} \modelpostcov \linmap^T \obspriorcov^{-1} \, \noise \, .
    \end{split}
\end{equation}
Since $\noise$ and $\obsnoise$ follow the same distribution, it is clear that
\begin{equation}
    \covrep = \covcent,
\end{equation}
which simply means that
\begin{equation}
    \var = \mathbf{E}_{\obspriorcent}[{\rm bias}].
\end{equation}
We recall that when the map is linear, the NNPDF MC methodology generates
replicas which are sampled from the posterior distribution of the model given
the data. We have shown here that provided the underlying law belongs to the
model space, the posterior distribution of the model predictions satisfy the
requirement that $\biasvarratio = 1$.

We note that due to the invariance of the trace under cyclic permutations, we
can rearrange Eq.~\ref{eq:BiasLinearModel} as
\begin{equation}
    \mathbf{E}_{\obspriorcent}[{\rm bias}] = \frac{1}{\ndata}
    {\rm Tr} \left[
        \modelpostcov
        \testset{\linmap}^T \testset{\obspriorcov}^{-1} \testset{\linmap}
    \right] \, ,
\end{equation}
where the term $\testset{\linmap}^T \testset{\obspriorcov}^{-1}
\testset{\linmap}$ can be understood as the covariance matrix of the posterior
distribution in model space given the test data, with zero prior knowledge of
the model, which we denote as $\modelpostcov'$:
\begin{equation}\label{eq:BiasTraceModelPost}
    \mathbf{E}_{\obspriorcent}[{\rm bias}] = \frac{1}{\ndata}
    {\rm Tr} \left[ \modelpostcov {\modelpostcov}^{ \prime -1} \right] \, ,
\end{equation}
where we emphasise that the covariance matrices $\modelpostcov$ and
$\testset{\modelpostcov}$ are obtained from completely independent Bayesian
inferences with no prior information on the model parameters, unlike in
Eq.~\ref{eq:ModelPostSequential} where a sequential marginalisation causes
$\testset{\modelpostcov}$ to depend on $\modelpostcov$.

Alternatively, if we perform a sequential marginalisation of the data, and use
the result in Eq.~\ref{eq:ModelPostSequential}, but then take
$\testset{\obspriorcov}^{-1} \to 0$, \ie\ there is no information on the
observables in the test set, then the covariance of the posterior model
distribution is 
\begin{equation}
    \modelpostcov^{-1} = \linmap^T \obspriorcov^{-1} \linmap \, ,
\end{equation}
which is identical to the posterior model distribution given just the training
data - as one would expect. Now we can express bias (or variance) as
\begin{equation}\label{eq:BiasTraceObsPost}
    \mathbf{E}_{\obspriorcent}[{\rm bias}] = \frac{1}{\ndata}
    {\rm Tr} \left[
        \testset{\obspostcov}
        \testset{\obspriorcov}^{-1}
    \right] \, ,
\end{equation}
where $\testset{\obspostcov}$ is the covariance of the posterior distribution of
$\testset{\obs}$ with no prior information on that data. This might seem
peculiar because in determining $\testset{\obspostcov}$ we took the limit
$\testset{\obspriorcov}^{-1} \to 0$, which encodes the fact that we had no prior
information on the unseen data, however in Eq.~\ref{eq:BiasTraceObsPost} we
require $\testset{\obspriorcov}^{-1}$ to be finite. We rationalise
Eq.~\ref{eq:BiasTraceObsPost} as a comparison between the posterior distribution
in the space of data of some unseen observables to an independently determined
prior from performing the relevant experiment which measures the same
observables. The two distributions can be compared when the independently
measured experimental data is published.

\paragraph{Underparameterised model}

Note that if we were to choose the number of model parameters such that $\nlaw >
\nmodel$, then the variance would be unaffected, since the underlying law
parameters cancel. However, the bias would now contain an extra term from the
extra parameters in the underlying law, schematically:
\begin{equation}
    \begin{split}
        (\emodel{\testset{\fwdobsop}\left(\modelvecrep\right)} - \testset{\law})_i =
        \sum_{1 \leq j \leq \nmodel} \testset{\linmap}_{ij} (\modelpostcent - \lawmodel)_j -
        \sum_{\nmodel < j \leq \nlaw} \testset{\linmap}_{ij} \lawmodel_j,
    \end{split}
\end{equation}
which would mean that $\biasvarratio \neq 1$. This demonstrates that requiring
$\biasvarratio = 1$ demands that the model space is suitably flexible, if the
underlying law is parameterised then this can be summarised as requiring
$\lawmodel \in \modelspace$. Note that in the underparameterised regime the
model replicas are still drawn from the posterior distribution, however because
$\lawmodel \notin \modelspace$ we have somehow invalidated the assumptions that
go into the relation between model predictions and the data-space prior.

Although $\biasvarratio$ was largely chosen on practical grounds, we see that it
is still a stringent test that our assumptions are correct and that the
distribution our model replicas are drawn from is meaningful, this is what we
mean when we say {\em faithful uncertainties}.

An unfortunate trade-off when using $\biasvarratio$ is that it cannot be used as
a diagnostic tool, and is instead used simply for validation. For example, if
$\biasvarratio > 1$, then we cannot know whether there was a problem with the
fitting methodology used to generate the model replicas or a deeper issue such
as an inflexible model.

%% file: results.tex
\section{Numerical Setup and Results}
\label{sec:numerical-results}

In this section we first introduce the experimental setup used to run the
closure tests, and then discuss the actual results. Following the study
performed in Ref.~\cite{nnpdf30}, we first analyse the relative size of the
different components of PDFs uncertainty, comparing the changes between methodologies
used to produce the NNPDF3.1 \cite{Ball_2017} and
NNPDF4.0 \cite{NNPDF40} sets of PDFs respectively. We then move to the data
space estimators $\biasvarratio$ and $\xisigdat{1}$, which have been computed
only for NNPDF4.0 \footnote{As pointed out before, the computation of the
expectation value over training data, defined in
Eq.~\ref{eq:average_over_training_data}, is made possible by the efficiency of
the new framework \cite{nnpdf40code}.}. The results here act both as a proof of
principle of new estimators presented in this paper but also as part of a suite
of methodological validation tools, see also the "future tests"
\cite{Cruz_Martinez_2021}, used to understand the PDF uncertainties of the
recent NNPDF4.0 set of PDFs. For the purpose of understanding how the results
here were produced, we will briefly describe the key features of the NNPDF4.0
methodology, but refer the reader to Ref.~\cite{NNPDF40} for a full discussion on how these
methodological choices were made.

\subsection{Closure test setup}

Using neural networks to fit PDFs has been discussed many times in previous
NNPDF publications, see for example \cite{nnpdf30, Ball_2017}. A new feature of
NNPDF4.0 is that, for the default fit performed in the evolution basis, a single
neural network parameterises all 8 PDF flavours $\{ g, \Sigma, V, V_3, V_8, T_3,
T_8, T_{15} \}$ at the initial scale. The PDF for a single flavour $j$, at the
initial scale $Q_0 = 1.65~{\rm GeV}$ is given by
\begin{equation}
    f_j(x, Q_0) = NN(x, \ln x | \modelvec)_j * x^{1-\alpha_j} * (1-x)^{\beta_j},
\end{equation}
where $\alpha$ and $\beta$ are the preprocessing exponents, which control the
PDF behaviour at $x \to 0$ and $x \to 1$ respectively and $NN(x, \ln x |
\modelvec)_j$ is the $j^{\rm th}$ output of the neural network, which takes $x$
and $\ln x$ as input. As discussed in Sec.~\ref{sec:fit-reps}, an ensemble of
models is fitted, each one is a MAP estimator of the corresponding pseudo-data
it is fitted on. An optimization algorithm is used to try and find the
parameters which maximise the likelihood. 
There are clearly many choices with respect to
hyperparameters, the discussion of how these choices have been made is beyond
the scope of this paper and left to the full NNPDF4.0 release \cite{NNPDF40}. A
summary of the hyperparameters used to produce results presented in this paper
are provided in Tab.~\ref{tab:Hyperparams}.

\begin{figure}
    \centering
    \includegraphics[width=0.8 \textwidth]{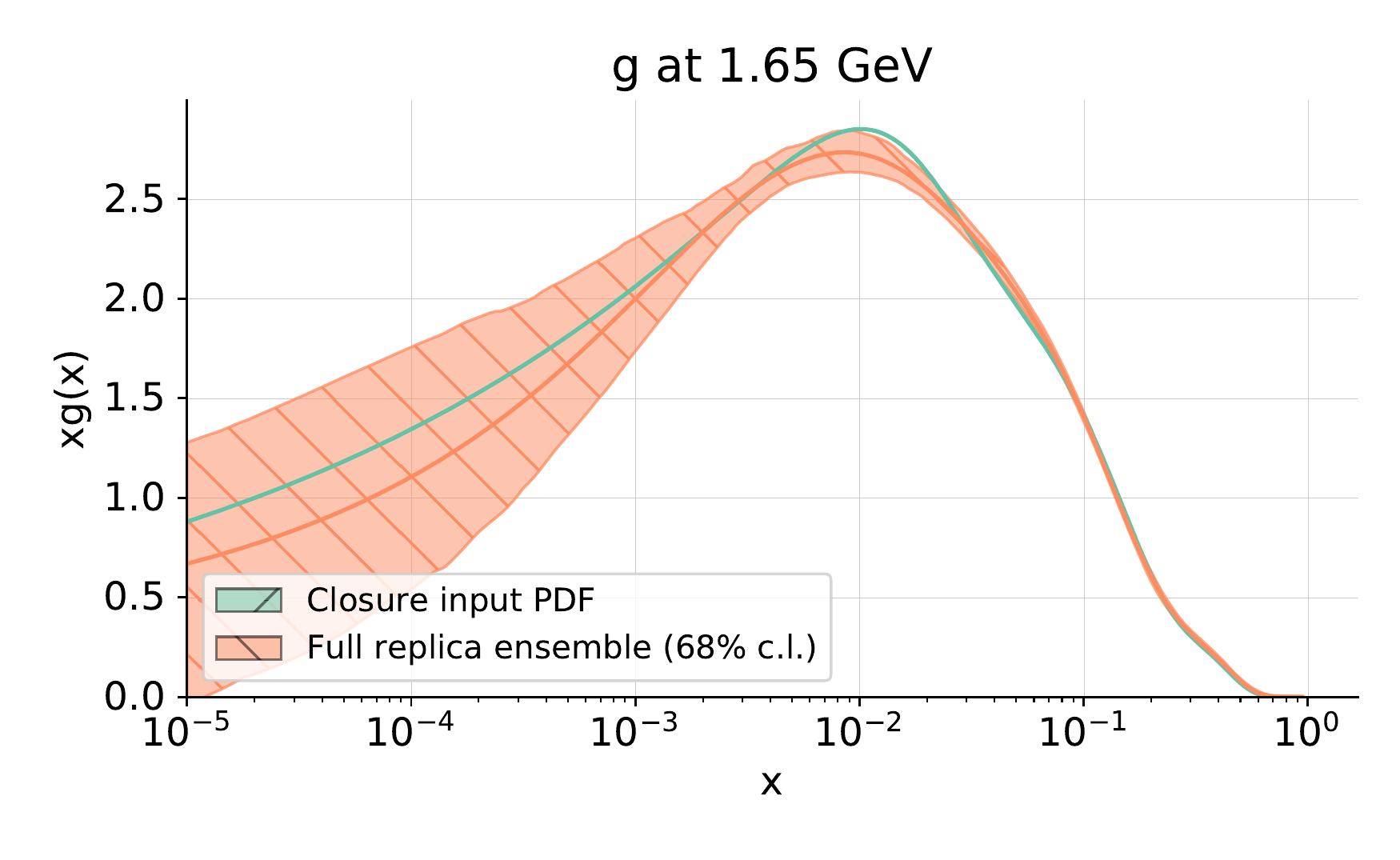}
    \caption{The green line is the input underlying law for the gluon PDF, which
    is sampled from the ensemble of replicas from a fit to data. The central
    value and the 68\% confidence interval for those replicas are plotted as the
    orange line and band respectively.}
    \label{fig:InputGluonPDF}
\end{figure}

As input to the closure test, a single replica was drawn randomly from a
previous NNPDF fit to experimental data. This
has generally more structure than the final central PDF and it is therefore a
more general choice than any final central fit. We refer to this as the underlying law
and the corresponding predictions as the true observable values. An example of
the gluon input is provided in Fig.~\ref{fig:InputGluonPDF}. In principle any
function could be used as underlying law, however it makes sense to use a
realistic input. 

The observables used in the fits are a subset of the full NNPDF4.0 dataset. For
convenience, we chose to fit the PDFs on a variant of the NNPDF3.1 dataset used
in Ref.~\cite{Ball_2018}, which is described in detail in a study of the
determination of the strange PDF~\cite{Faura_2020}. The datasets used in the
calculation of statistical estimators are the new datasets which have been
included in NNPDF4.0 and discussed in detail in the main release \cite{NNPDF40}. For a
full summary of the observables used in the test data and a visual representation of
the kinematic region of both the training and testing data, see
App.~\ref{sec:appendix-datasets}.

The partitioning of the available data into fitting and testing should not
affect the interpretation of the closure test results. However, one
could consider splitting the data into fitting and testing in a way which was
physically motivated \eg\ the partitioning could have been
stratified such that the kinematic coverage of the data of each partition
was approximately equal.
Alternatively, since the data is generated from the theory predictions produced
by the input underlying law, one could even produce completely artificial data
using a different set of FK tables. 

In order to compute the expectation value over training data defined in
Eq.~\ref{eq:average_over_training_data}, we generate 30 different sets of
experimental central values (or L1 data), as discussed in
Sec.~\ref{sec:closure-test-intro}, for the fitted 3.1-like dataset. Each set of
experimental central values was then fitted following NNPDF4.0 methodology
\cite{NNPDF40}, producing 40 pseudo-data replicas.

\subsection{Different components of the PDF uncertainty}

As already discussed in Ref.~\cite{nnpdf30}, fitting to L0, L1 and L2
pseudo-data allows us to validate different aspects of the fitting procedure. In
an L0 fit, we fit multiple time the exact same set of data, which corresponds to
the theory prediction from the chosen model. The fitted pseudo-data is the
result obtained by applying the forward map to the model. It is clear that in
this case the quality of the fit can be improved at will, provided that the
parametrization is flexible enough and the minimization algorithm is efficient.
There are indeed multiple solutions that reproduce exactly the data set, while
interpolating between the data points. In an L1 fit the data have been
fluctuated around the theoretical prediction -- mimicking thereby the central
values of experimental measurements. The true model no longer reproduces the
data; instead it will yield a $\chi^2$ of order one. The pseudo-data are held
fixed, fluctuations from one replica to the next are due to the existence of
multiple solutions that hold a similar value for the residual $\chi^2$ at the
end of the minimization process. Finally, in the L2 fits, the fluctuations of
the data are reproduced by the replicas, and propagated to the model function
when fitting the data for each replica. Since the NNPDF fitting methodology has
changed in the latest release, adopting the procedure described in
Ref.~\cite{Carrazza:2019mzf}, it is important to compare the uncertainties at
L0, L1 and L2 that are obtained with the NNPDF4.0 methodology with the
corresponding ones obtained with the NNPDF3.1 framework.

\begin{figure}[h]
    \centering
    \includegraphics[scale=0.43]{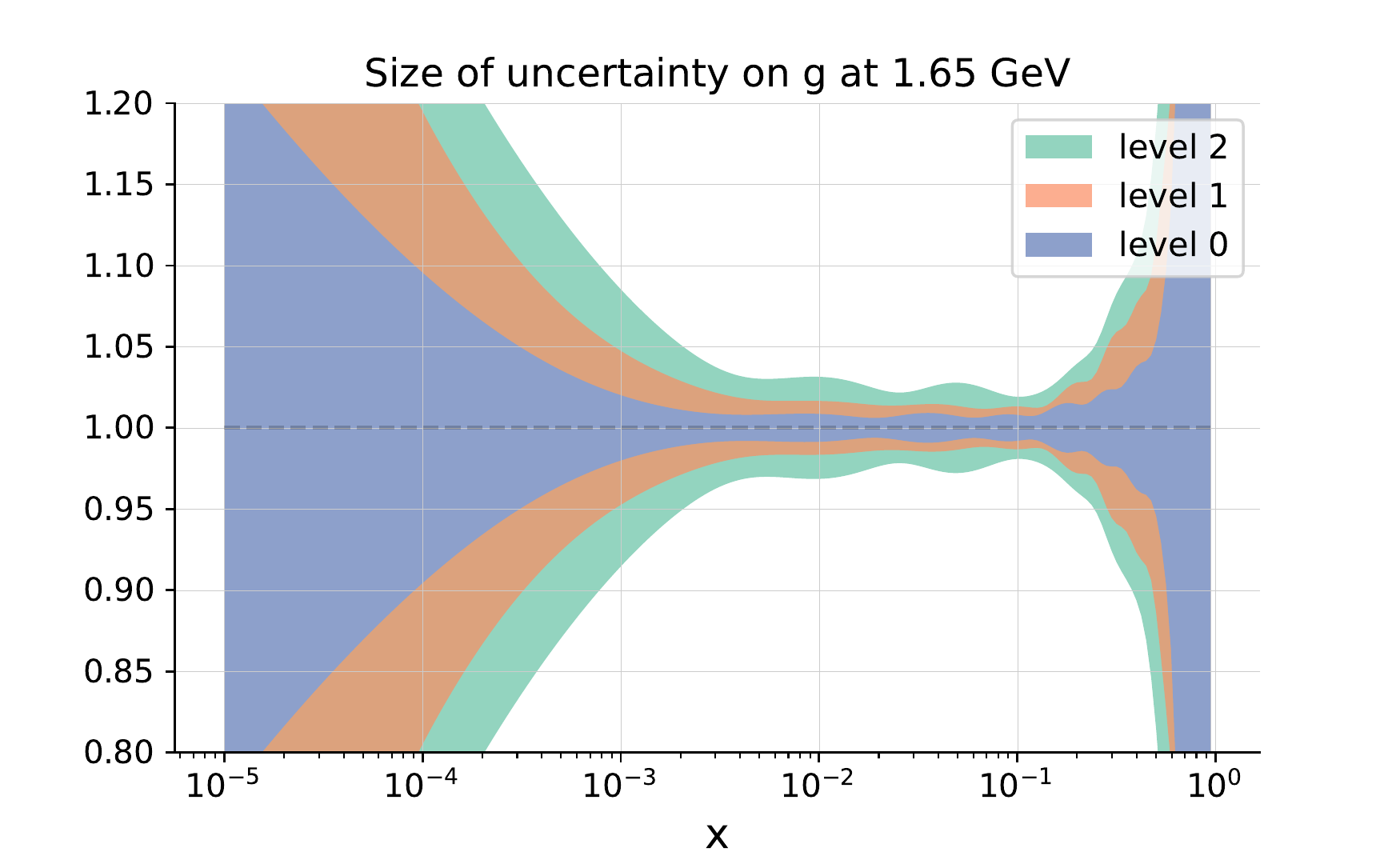}
    \includegraphics[scale=0.43]{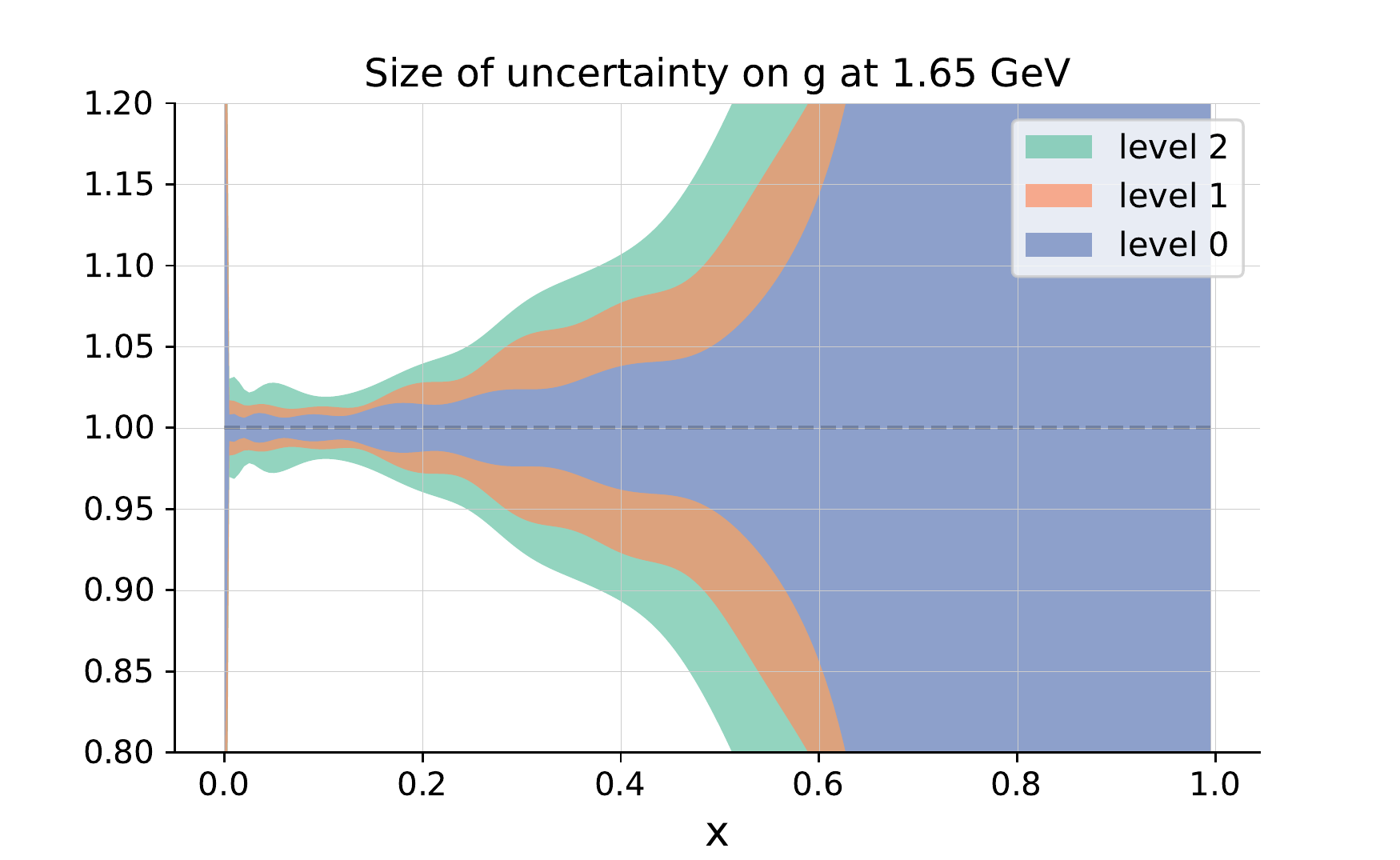}
    \includegraphics[scale=0.43]{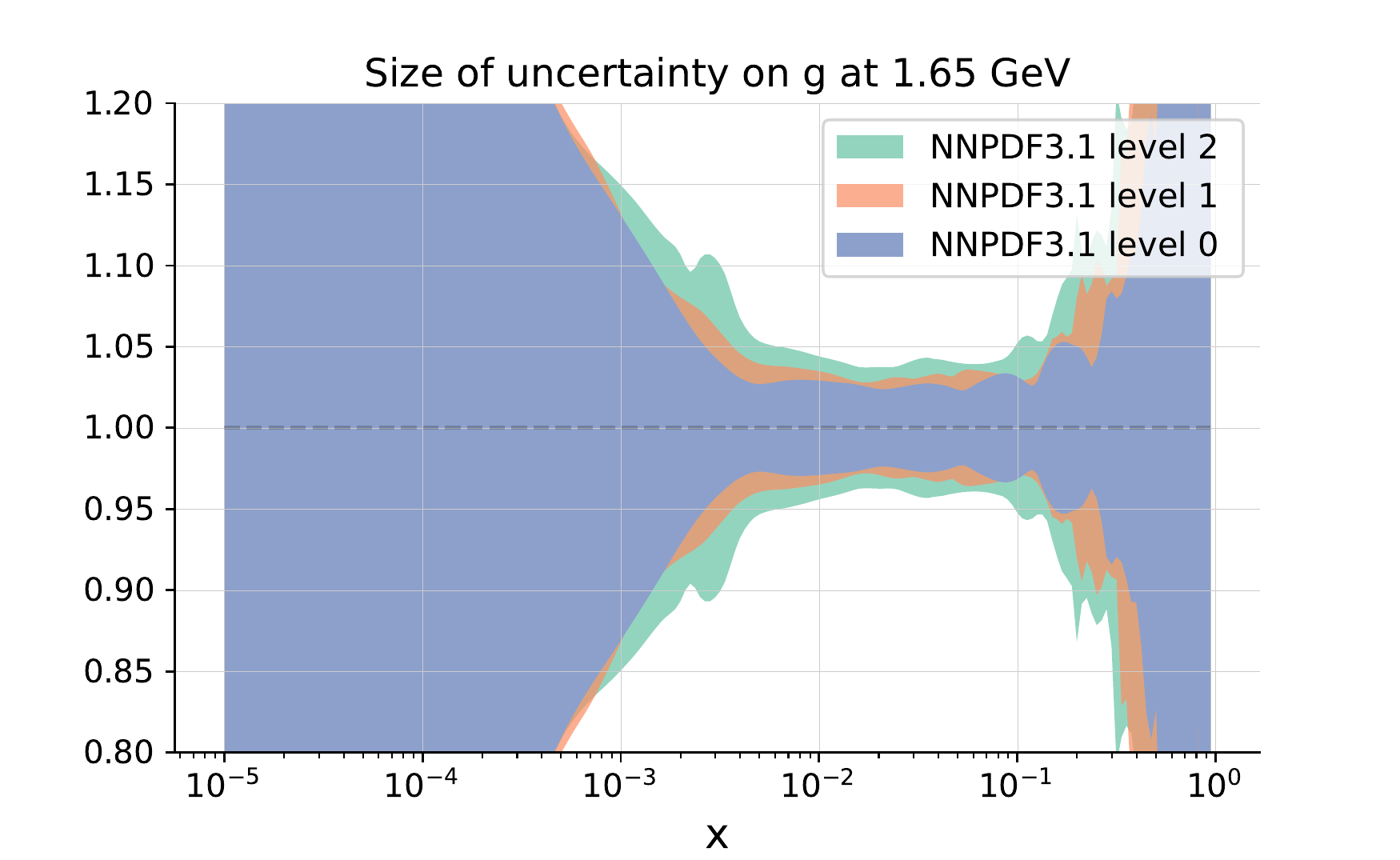}
    \includegraphics[scale=0.43]{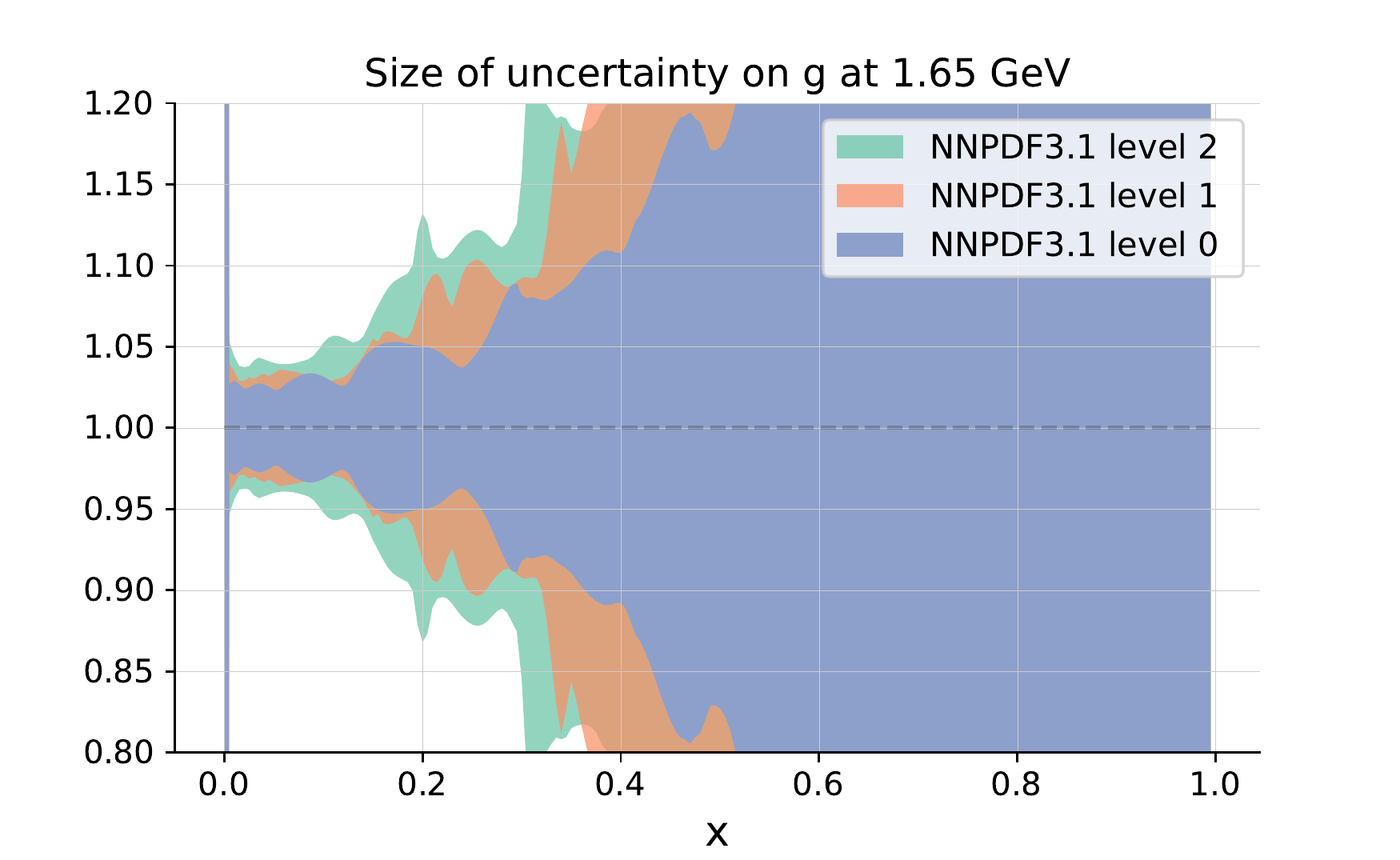}
    \caption{Relative PDF error for the gluon distribution in the NNPDF4.0 (upper panel)
    and NNPDF3.1 (lower panel) methodology, plotted in logarithmic (left) and linear scale (right).}
    \label{fig:CT_uncertainty_g}    
\end{figure}

An example of these relative errors on fitted PDFs is shown in the upper panel of
Fig.~\ref{fig:CT_uncertainty_g} where results from L0, L1 and L2 closure tests
are displayed on the same plot in the case of the gluon distribution. Each fit
is normalized to the corresponding central value. We note how the L0 and L1
uncertainties tend to increase in the $x$ regions where less experimental data
are available, namely at small and large-$x$, where the model is left
unconstrained and has more freedom to fluctuate, while they are considerably
reduced in the data region where the contribution of the L2 error, induced by
the actual experimental data, becomes more important. The fact that the data
uncertainty is not always the dominant component of the PDF uncertainty was
already stressed in Ref.~\cite{nnpdf30}. Improved methodologies should therefore
aim to reduce the L0 and L1 error. 

\begin{figure}[h]
    \centering
    \includegraphics[scale=0.43]{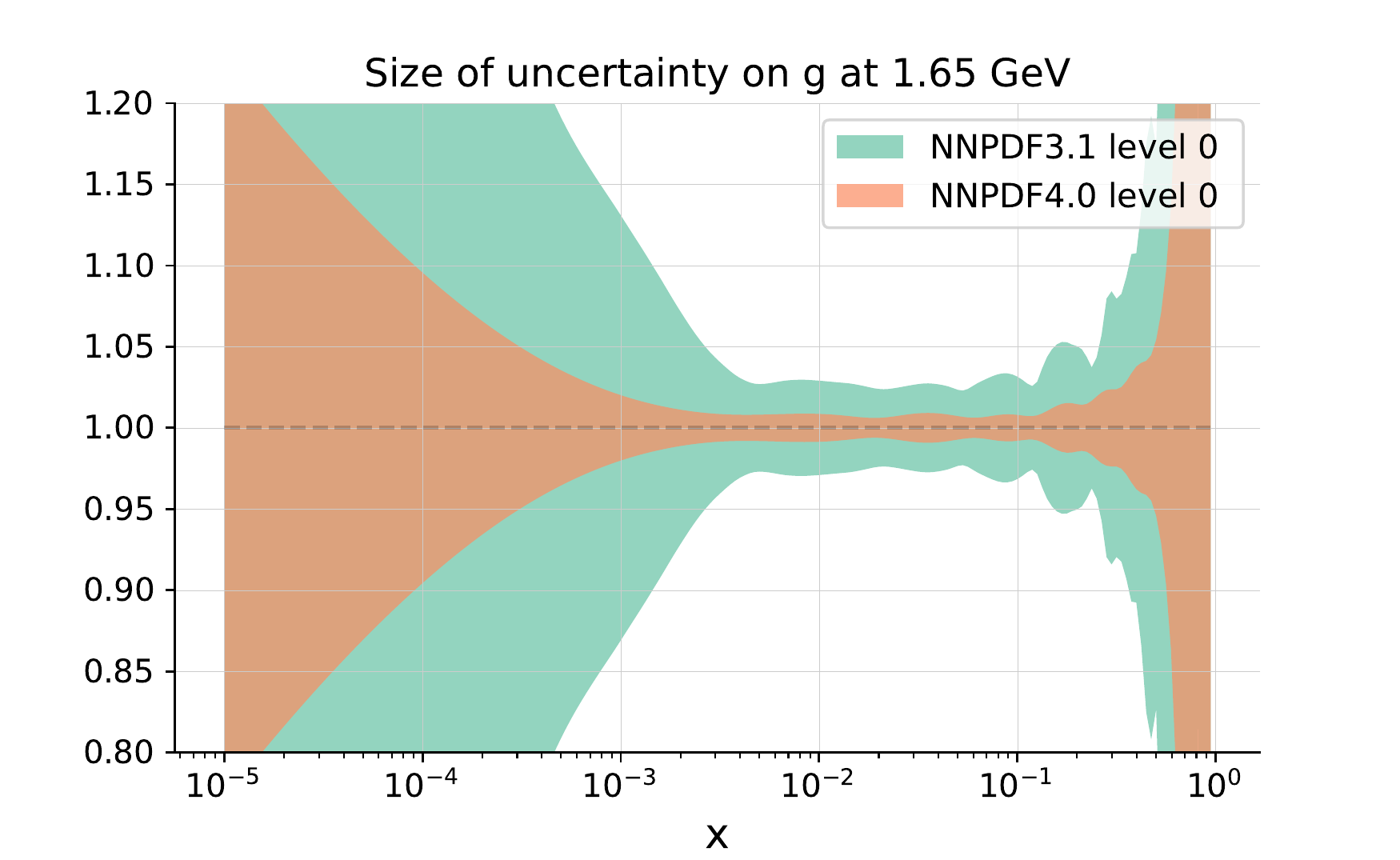}
    \includegraphics[scale=0.43]{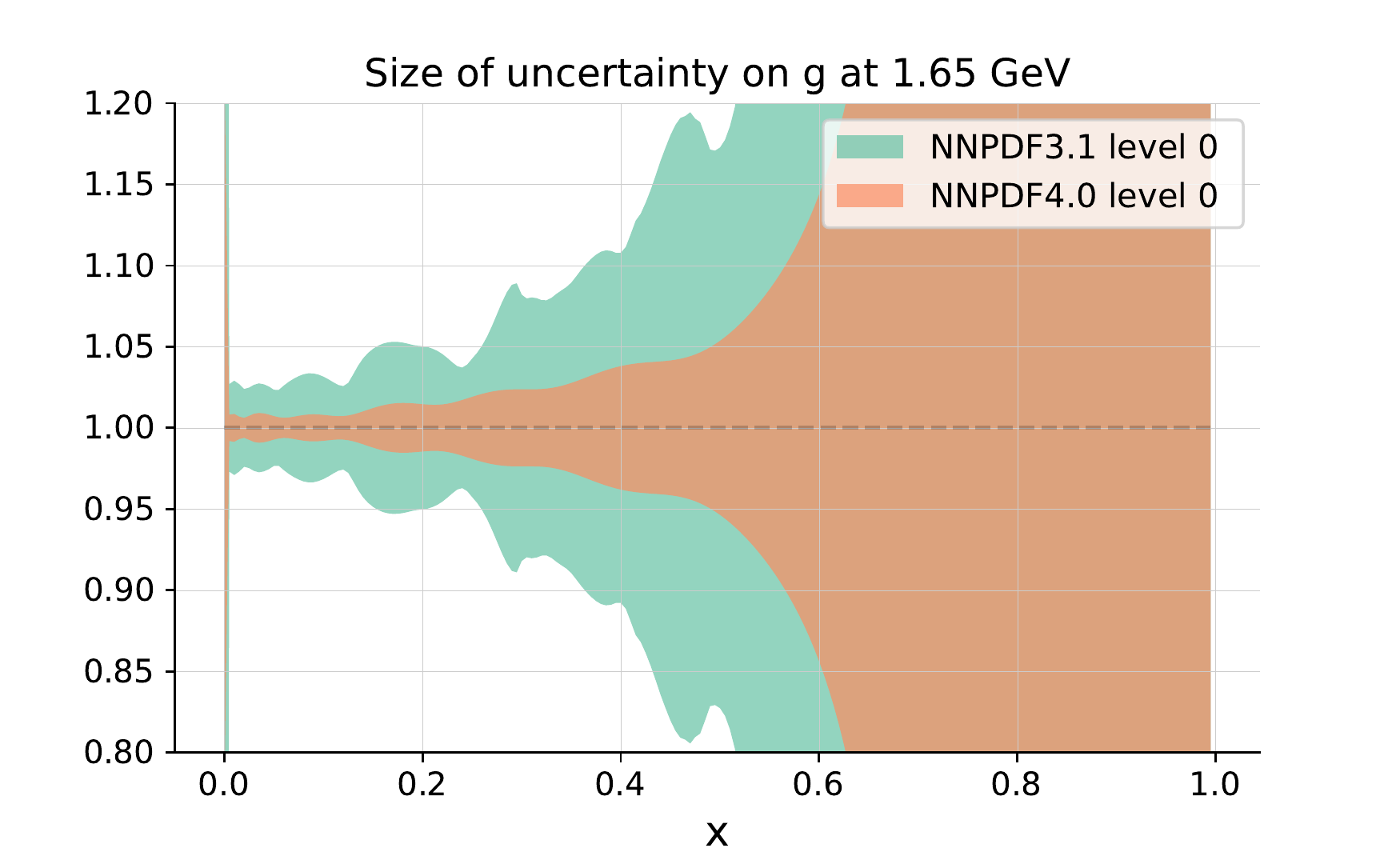}
    \caption{Level 0 uncertainty in the NNPDF4.0 and NNPDF3.1 methodology.}
    \label{fig:CT_uncertainty_new_vs_old}    
\end{figure}

\begin{figure}[h]
    \centering
    \includegraphics[scale=0.43]{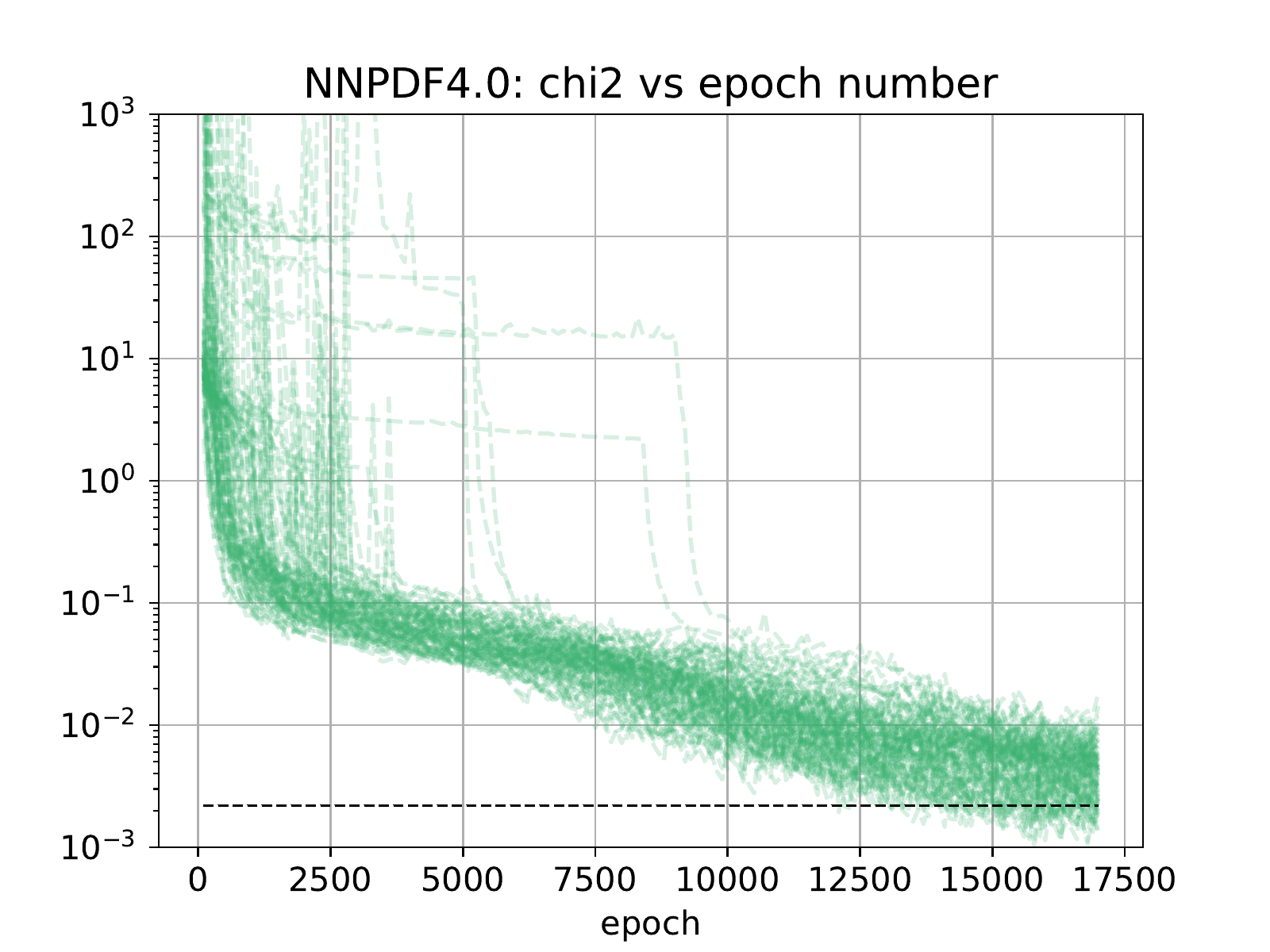}
    \includegraphics[scale=0.43]{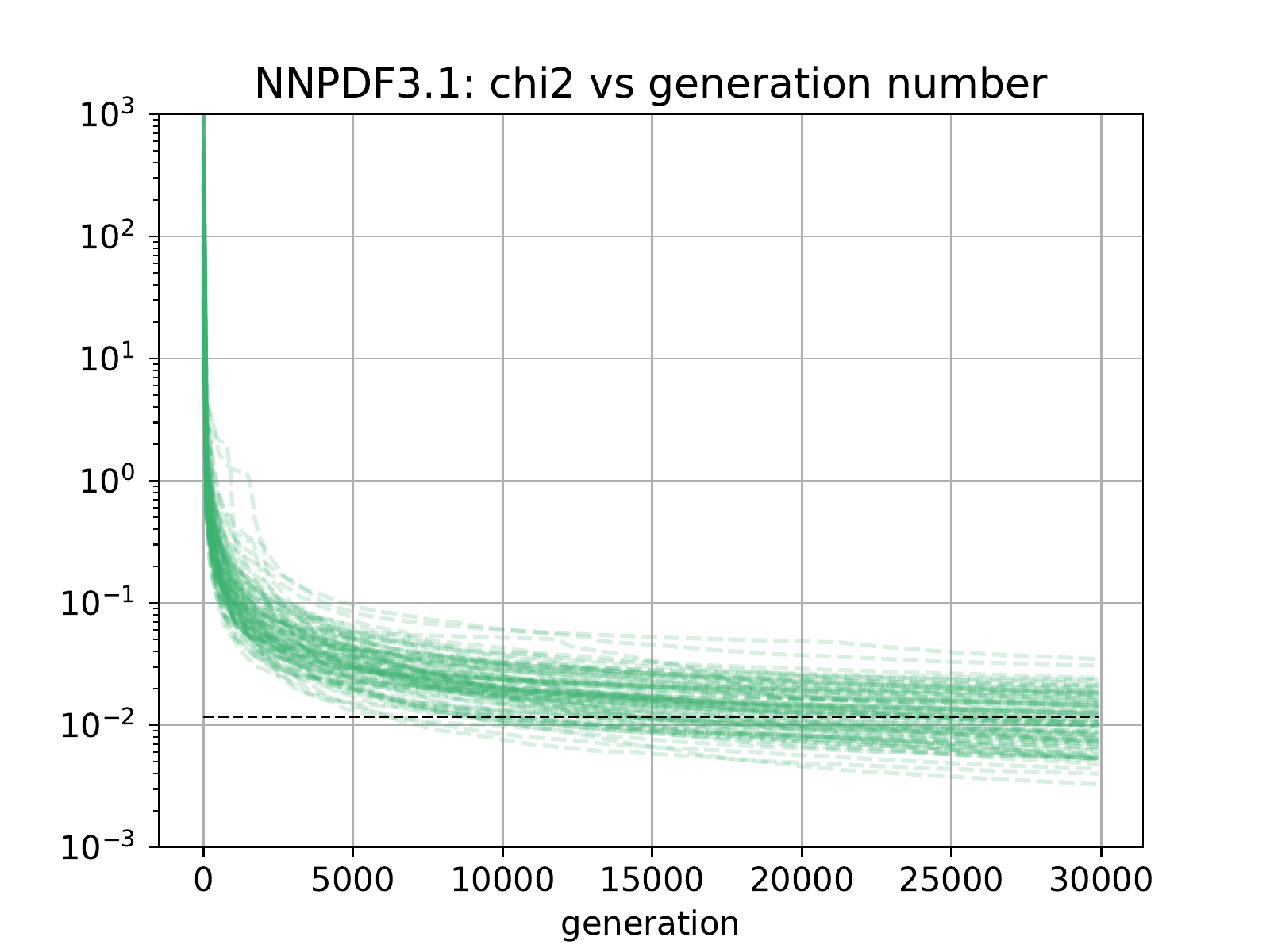}
    \caption{$\chi^2$ distribution across replicas as a function of the training
    epoch and genetic algorithm generation for the NNPDF4.0 (left) and NNPDF3.1
    (right) methodology. The black dotted line
    in each plot represents the final $\chi^2$ of the central value of the
    corresponding fit, equal to $0.002$ and $0.012$ respectively.}
    \label{fig:chi2_vs_epoch}    
\end{figure}

In this respect, it is interesting to compare these results with what we find
with the NNPDF3.1 methodology. The corresponding plots are shown in the lower panel of
Fig.~\ref{fig:CT_uncertainty_g}: unlike the NNPDF4.0 case, here the PDF
uncertainty is always dominated by the L0 uncertainty, even in those kinematic
regions where experimental data are present. We can conclude that moving to the
NNPDF4.0 methodology, thanks to the optimized hyperparameters listed in Tab.~\ref{tab:Hyperparams},
we observe a marked reduction of the L0 uncertainty, 
see Fig.~\ref{fig:CT_uncertainty_new_vs_old}, where the latter is plotted on the same panel
for both NNPDF4.0 and NNPDF3.1. 
The better efficiency of the NNPDF4.0 methodology can be also appreciated
by looking at Fig.~\ref{fig:chi2_vs_epoch}, where we plot the distribution
across replicas of the L0 closure test $\chi^2$
as a function of the training epoch and genetic algorithm generation
for the NNPDF4.0 and NNPDF3.1 methodology respectively. As mentioned before, in a L0
closure test the quality of the fit can in principle be improved at will,
assuming the use of an efficient optimizer. For the final $\chi^2$ of the
central value of each fit, plotted as a black dashed line, we find $0.002$ and
$0.012$ in the NNPDF4.0 and NNPDF3.1 fitting code respectively, showing the better
efficiency of the new methodology. 

It should be noted how
Fig.~\ref{fig:CT_uncertainty_g}  only
provide a qualitative assessment of the relative size of the different
components of PDFs uncertainty. Despite being useful to assess how the
methodology has improved with respect to the previous one, they do not provide
any quantitative estimation of the faithfulness of PDF uncertainty. This is best
achieved in data space, using the new estimators introduced in the previous
sections.

\subsection{Data space estimators}

Results for $\biasvarratio$ and the corresponding uncertainty calculated on the
test data are shown in the first column of Tab.~\ref{tab:biasvarratio}.
The uncertainties, which take into account the finite size of the replicas and fits samples,
have been computed by performing bootstrap, 
\cite{efron1994introduction}, where we randomly sample from both fits and
replicas and re-calculate $\biasvarratio$. The value and error presented in the
table is then the mean and standard deviation across bootstrap samples. We
checked that the distribution of the estimator across bootstrap samples is
indeed Gaussian. We also checked that increasing the number of fits and replicas
reduced the bootstrap error but the central values were the same within the
estimated bootstrap uncertainties. We see that overall $\biasvarratio$  is
consistent with 1, which gives a good indication that, at least for the unseen
data used in this study, the uncertainties are faithful

\begin{table}[h]
    \begin{center}
        \setlength{\tabcolsep}{12pt} 
        \begin{tabular}{rrr}
            \toprule
             $\biasvarratio$ & $\xi_{1\sigma}$ & $\erf(\biasvarratio/\sqrt{2})$ \\
            \midrule
             $1.03\pm0.05$ & $0.69\pm0.02$   & $0.67\pm0.03$                  \\
            \bottomrule
            \end{tabular}
    \end{center}
    \caption{In the first column we show the bias-variance ratio,
        $\biasvarratio$, for unseen data, summarised in
        Tab.~\ref{tab:summarise_new_data}. The uncertainty is estimated by
        performing a bootstrap sample across fits and replicas and calculating
        the standard deviation. We see that overall $\biasvarratio$ is
        consistent with 1, within uncertainties. In the second and third
        columns we compare the measured value of $\xi_1\sigma$ and the estimated
        value from $\biasvarratio$. The two values are consistent, which
        suggests the approximation that the ratio of uncertainties is
        approximately the same across all data is not completely invalidated.}
    \label{tab:biasvarratio}
\end{table}

\begin{figure}[h]
    \centering
    \includegraphics[width=0.6 \textwidth]{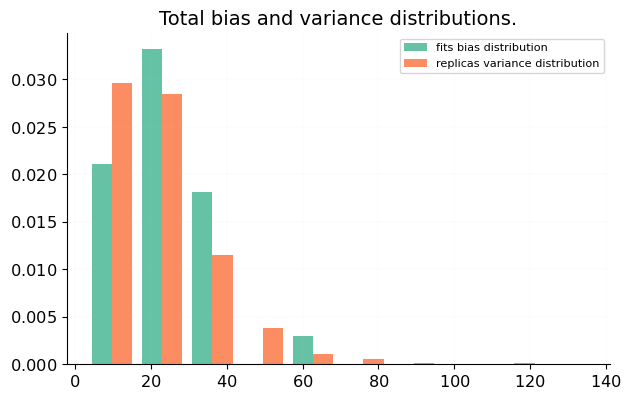}
    \caption{The green histogram is the distribution of the total bias across
    fits, the orange histogram is the distribution of the difference between the
    replica and central predictions squared, in units of the covariance across
    all fits and replicas. This gives a qualitative picture of the full
    distribution, in Tab.~\ref{tab:biasvarratio} we compare the square root of
    the mean of each distribution, getting the quoted value for
    $\biasvarratio$.}
    \label{eq:bias_varinace_distributions}
\end{figure}

In Fig.~\ref{eq:bias_varinace_distributions} we compare qualitatively the
distribution of bias across fits, to the distribution of the difference between
replica predictions and expectation values of predictions (in units of the
covariance) across different fits and replicas. The square root ratio of the
mean of these two distributions is precisely $\biasvarratio$.

As discussed in Sec.~\ref{sec:QuantileStatistics}, one can define an analogous
estimator in data space, based upon $\xi_{n\sigma}$, which was defined on a grid
of points in $x$ and $Q^2$ in PDF space in \cite{nnpdf30}. There is not a
one-to-one correspondence between this and $\biasvarratio$, but a loose
approximation using Eq.~\ref{eq:expectedxi}. In the second and third columns of
Tab.~\ref{tab:biasvarratio} we compare the estimated $\xi_{1\sigma}$ from
substituting $\biasvarratio$ into Eq.~\ref{eq:expectedxi} and to the measured
value. Despite the assumptions entering each of the two estimators differing, we
see good agreement between the $\xi_{1\sigma}$ estimated from $\biasvarratio$
and that measured directly. We find this result reassuring, since it indicates
not only that the total uncertainty averaged across all data is faithful, but
also that the uncertainty on each data point seems faithful. If the results
differed it would indicate some kind of imbalance, where some components of the
uncertainty are correctly represented by the replicas but other directions are
not. Finally we note how not only are the measured value and estimated value
from $\biasvarratio$ self consistent, but they are also consistent with $0.68$,
which further supports the argument that the model uncertainties are faithful.

%% file: summary.tex
\section{Summary}

We have presented a theoretical framework for treating inverse problems from a
Bayesian perspective. In particular, the framework provides a more formal
description of what it means when we talk about propagating experimental
uncertainties into the space of models by focusing on the posterior probability
measure in model space given the available data. Strictly speaking, there is no
fitting required to obtain the expression for the posterior distributions in the
space of the data or the model, instead these are obtained by marginalising the
joint distribution. We note that sampling from the posterior distribution of the
model is, in general, highly non-trivial; however we can show that at least for
linear problems the NNPDF MC methodology produces a sample of models which are
distributed according to the posterior model distribution predicted from the
Bayesian formalism. Furthermore, we provide evidence that even for non-linear
models this result at least holds as a good approximation close to the MAP
estimator.

We then use this formal framework to think about some of the estimators which we
use as part of the NNPDF closure test. In particular we derive bias and variance
from decomposing an out of sample error function, which is understood from a
classic fitting point of view. The estimators are then related back to the
posterior distributions in the Bayesian framework. We note that the estimators
themselves are not perfect and suffer from only testing the model uncertainties
locally (in regions where the test data probes). Furthermore, the estimators
only give an approximate overall picture, and cannot be used to diagnose where
the problem arises if we find evidence that the model uncertaintes are not
faithful.

Given the framework set out here, future work should be undertaken to generalise
the closure estimators to model space. This would likely involve a combination
of the closure estimators presented here and the extension of the Bayesian
framework to infinite-dimensional spaces.

We give some closure results, as a proof of principle. The results
presented here are examined in more detail alongside the full NNPDF4.0
release \cite{NNPDF40} but serve as an example of how the data space
estimators can be practically included even in a rather complex setting. 
The NNPDF4.0 methodology
passes the closure test according to the estimators that we analysed, providing
evidence that for unseen data the current NNPDF methodology appears to provide
faithful uncertainties. The estimators are not limited to this specific
application, and the results presented demonstrate how the data space estimators
can be incorporated into any inverse problem. As previously mentioned, a more
general set of estimators in model space would be the gold-standard and give us
confidence in our uncertainties for future observables which probe regions which
are not covered by either the training or testing data.

In the closure test framework described in this study, the prior on the data is
fully consistent with the generation of the observable central values from the
true observables and uncertainties by construction, which is likely not the case
in real world fits. Something which should be investigated is how we can
guarantee faithful uncertainties when this is no longer the case. For example
one could consider a closure test where the generated data was inconsistent with
the prior. Most observed inconsistentency between the prior and the data central
values is likely due to missing theoretical uncertainties, but once all sources
of theoretical uncertainties have been accounted for there could still be
tension in the data. The advantage of viewing inverse problems from a Bayesian
perspective is access to methodologies which deal with inconsistent data (or
unknown systematics) in a Bayesian framework, for example in these cosmological
studies~\cite{Luis_Bernal_2018, Hobson_2002}. These methods potentially offer a
more formal approach to dealing with inconsistent data, rather than ad hoc
procedures.

The inclusion of MHOUs in the likelihood was justified from a Bayesian
perspective~\cite{AbdulKhalek:2019ihb}, but here we have drawn a connection
between the model replicas and the posterior distribution in model space, which
explains why theoretical uncertainties must be included in both the sampling of
the pseudodata replicas and the likelihood. Therefore we emphasise that the
framework set out here is not only useful for understanding model uncertainties
with the current methodology, but also for motivating future methodological
development from a Bayesian perspective. We defer more detailed investigations
of the potential of this approach to future work. 

\section*{Acknowledgements}
We thank the members of the NNPDF collaboration for discussions and comments on the present work. 
L.~D.~D. is supported by the U.K. Science and Technology Facility Council
(STFC) grant ST/P000630/1, T.~G. is supported by NWO via a ENW-KLEIN-2 project and
M.~W. by the STFC grant ST/R504737/1.

%% file: gaussian.tex
\section{Gaussian integrals}
\label{sec:GaussianIntegrals}

Theory errors can be included in this framework by allowing the distribution of
observables around the theory prediction to have a finite width, \eg\ by
replacing the Dirac delta 
\begin{equation}
    \label{eq:DiracInApp}
    \delta(y-\mathcal{G}u)
\end{equation}
in Eq.~\ref{eq:ThetaCorr} with a Gaussian
\begin{equation}
    \label{eq:TheoryGaussian}
    \theta(\obs,\modelvec|\fwdobsop) \propto \exp\left[
        -\frac12 \left(y-\mathcal{G}u\right)^T
        C_T^{-1} \left(y-\mathcal{G}u\right)
    \right]\, .
\end{equation}
For the purposes of this study, we do not want to provide a realistic estimate
of theory errors. Instead we will be assuming that the errors are uncorrelated
and identical for all data points
\begin{equation}
    \label{eq:DiagTheoryCov}
    C_T = \sigma^2 \mathds{1}\, ,
\end{equation}
and we will be interested in the limit where $\sigma^2\to 0$. 

\subsection{Integrating out the data}
\label{sec:IntOutData}

Marginalizing with respect to \obs\ in this case yields 
\begin{align}
  \label{eq:MarginGaussData}
  \pi_M(\modelvec|\obspriorcent,\modelpriorcent,\fwdobsop) 
  &\propto \pi_{M}^0(\modelvec|\modelpriorcent) \, 
  \int dy\, \pi_{D}^0(\obs|\obspriorcent) 
    \theta(\obs,\modelvec|\fwdobsop) \, .
\end{align}
The argument of the exponential in the integrand is a quadratic form in \obs, 
\begin{equation}
    \label{eq:QuadFormDataInt}
    A = \left(y-y_0\right)^T C_D^{-1} \left(y-y_0\right) +
    \left(y-\mathcal{G}u\right)^T C_T^{-1} \left(y-\mathcal{G}u\right)\, .
\end{equation}
The integral can be easily evaluated by completing the square, 
\begin{equation}
    \label{eq:QuadFormDataIntSquare}
    A = \left(y-\tilde{y}\right)^T 
    \tilde{C}_D^{-1}
    \left(y-\tilde{y}\right) + R_D\, .
\end{equation}
Comparing Eqs.~\ref{eq:QuadFormDataInt} and~\ref{eq:QuadFormDataIntSquare} at order $y^2$ and \obs, yields
\begin{align}
    \tilde{C}_D^{-1} &= \frac{1}{\sigma^2}
    \left(\mathds{1} + \sigma^2 C_D^{-1}\right)\, , \\
    \tilde{y} &= \left(\mathds{1} + \sigma^2 C_D^{-1}\right)^{-1} 
    \left(
        \mathcal{G}u + \sigma^2 C_D^{-1} y_0
    \right)\, ,
\end{align}
and therefore
\begin{align}
    \tilde{y}^T \tilde{C}_D^{-1} \tilde{y}
    &= \frac{1}{\sigma^2} \left(\mathcal{G}u\right)^T
    \left(\mathds{1}+\sigma^2 C_D^{-1}\right)^{-1} \left(\mathcal{G}u\right) +
    y_0^T C_D^{-1} \left(\mathds{1}+\sigma^2 C_D^{-1}\right)^{-1} 
    \left(\mathcal{G}u\right) + \nonumber \\
    \label{eq:RemainderFromSquare}
    & \quad + \left(\mathcal{G}u\right)^T C_D^{-1} 
    \left(\mathds{1}+\sigma^2 C_D^{-1}\right)^{-1} y_0 + 
    \sigma^2 y_0^T C_D^{-1} \left(\mathds{1}+\sigma^2 C_D^{-1}\right)^{-1} 
    C_D^{-1} y_0\, .
\end{align}
Note that the four terms in the equation above are ordered in increasing powers
of $\sigma^2$ and ultimately we will be interested in the limit $\sigma^2\to 0$,
which reproduces the Dirac delta in $\theta(y,u)$. Plugging
Eq.~\ref{eq:RemainderFromSquare} in Eq.~\ref{eq:QuadFormDataIntSquare} and again
comparing to Eq.~\ref{eq:QuadFormDataInt}, we find
\begin{equation}
    \label{eq:RDBeforeLimit}  
    \begin{split}
    R_D 
    &= \frac{1}{\sigma^2} \left(\mathcal{G}u\right)^T 
    \left[
        \mathds{1} - \frac{1}{\mathds{1}+\sigma^2 C_D^{-1}} 
    \right]
    \left(\mathcal{G}u\right) - y_0^T C_D^{-1} \left(\mathcal{G}u\right)
    - \left(\mathcal{G}u\right)^T C_D^{-1} y_0 + \\ 
    & \quad + y_0^T C_D^{-1} y_0 + \mathcal{O}(\sigma^2)\, ,       
    \end{split} 
\end{equation}
Expanding for small values of $\sigma^2$ the terms of order $1/\sigma^2$ cancel;
keeping only finite terms in the limit $\sigma^2 \to 0$ we finally obtain
\begin{equation}
    \label{eq:RDAfterLimit}
    R_D = \left(\mathcal{G}u - y_0\right)^T C_D^{-1}
    \left(\mathcal{G}u - y_0\right)\, .
\end{equation}
This is exactly the result that we obtained earlier when 
\begin{equation}
    \label{eq:RemindTheta}
    \theta(y,u|\mathcal{G}) = \delta(y-\mathcal{G}u)\, .
\end{equation}
It should not come as a surprise since in the limit where $\sigma^2 \to 0$ the
Gaussian distribution that we chose to describe the fluctutations of the data
around the theory predictions reduces indeed to a Dirac delta. The posterior for
the model is exactly the one we computed in Sect.~\ref{sec:inverse-problems}. We
do not learn anything new from this exercise, but it is a useful warm-up for the
next example. The integral over \obs\ can now be performed easily, since it is
yet again a Gaussian integral.

\subsection{Integrating out the model}
\label{eq:IntModOut}

Using the same approach as above, we now want to marginalise with respect to the model in order to obtain the posterior distribution of the data: 
\begin{align}
    \label{eq:MarginGaussModel}
    \pi_D(\obs|\obspriorcent,\modelpriorcent,\fwdobsop) 
    &\propto \pi_{D}^0(\obs|\obspriorcent) \, 
    \int du\, \pi_{M}^0(\modelvec|\modelpriorcent) 
      \theta(\obs,\modelvec|\fwdobsop) \, .
  \end{align}
We follow exactly the same procedure outlined above, starting from the argument
of the exponential
\begin{equation}
    \label{eq:QuadFormModelInt}
    A = \left(u-u_0\right)^T C_M^{-1} \left(u-u_0\right) +
    \left(y-\mathcal{G}u\right)^T C_T^{-1} \left(y-\mathcal{G}u\right)\, ,
\end{equation}
we complete the square and rewrite it in the form
\begin{equation}
    \label{eq:QuadFormModelIntSquare}
    A = \left(u-\tilde{u}\right)^T 
    \tilde{C}_M^{-1}
    \left(u-\tilde{u}\right) + R_M\, .
\end{equation}
It can be readily checked that in this case
\begin{align}
    \tilde{C}_M^{-1} &= \frac{1}{\sigma^2}
    \left(\mathcal{G}^T \mathcal{G} + \sigma^2 C_D^{-1}\right)\, , \\
    \tilde{u} &= \left(\mathcal{G}^T \mathcal{G} + \sigma^2 C_M^{-1}\right)
    \left(
        \mathcal{G}^T y + \sigma^2 C_M^{-1} u_0
    \right)\, .
\end{align}
In order to evaluate $R_M$, we need
\begin{equation}
    \label{eq:UtildeUtildeTerm}
    \begin{split}
    \tilde{u}^T \tilde{C}_M^{-1} \tilde{u} 
    &= \frac{1}{\sigma^2} y^T \mathcal{G} 
    \left(\mathcal{G}^T \mathcal{G} + \sigma^2 C_M^{-1}\right)^{-1}
    \mathcal{G}^T y +
    u_0^T C_M^{-1} 
    \left(\mathcal{G}^T \mathcal{G} + \sigma^2 C_M^{-1}\right)^{-1}
    \mathcal{G}^T y + \\
    & \quad + y^T \mathcal{G} \left(\mathcal{G}^T \mathcal{G} + \sigma^2 C_M^{-1}\right)^{-1} C_M^{-1} u_0 + \mathcal{O}(\sigma^2)\, .
    \end{split} 
\end{equation}
Noting that
\begin{equation}
    \label{eq:InverseFromTarantola}
    \left(\mathcal{G}^T \mathcal{G} + \sigma^2 C_M^{-1}\right)^{-1} =
    \frac{1}{\sigma^2} C_M - 
    \frac{1}{\sigma^2} C_M \mathcal{G}^T 
    \left(\mathcal{G} \frac{1}{\sigma^2} C_M \mathcal{G}^T 
        + \mathds{1}\right)^{-1} \mathcal{G}
    \frac{1}{\sigma^2} C_M\, ,
\end{equation}
we have, in the limit where $\sigma^2 \to 0$
\begin{equation}
    \label{eq:QuadraticYTerm}
    \mathcal{G} \left(\mathcal{G}^T \mathcal{G} + \sigma^2 C_M^{-1}\right)^{-1} 
    \mathcal{G}^T = 
    \mathds{1} - \sigma^2 \left(\mathcal{G} C_M \mathcal{G}^T\right)^{-1} + 
    \mathcal{O}(\sigma^4)\, .
\end{equation}
Collecting all terms we find
\begin{equation}
    \label{eq:RMAfterLimit}
    R_M = \left(y - \mathcal{G} u_0\right)^T 
        \left(\mathcal{G} C_M \mathcal{G}^T\right)^{-1}
        \left(y- \mathcal{G} u_0\right)\, .
\end{equation}
Performing the Gaussian integral over $u$ in Eq.~\ref{eq:MarginGaussModel}, we
obtain the posterior distribution of the data
\begin{equation}
    \label{eq:PosteriorDataDistr}
    \pi_D^y(y) \propto 
    \exp\left[-\frac12 \left(y-y_0\right)^T C_D^{-1} \left(y - y_0\right)
    -\frac12 \left(y - \mathcal{G} u_0\right)^T 
    \left(\mathcal{G} C_M \mathcal{G}^T\right)^{-1}
    \left(y - \mathcal{G} u_0\right)
    \right]\, .
\end{equation}
As in the case above, we note that this is a Gaussian distribution,
\begin{equation}
    \label{eq:PosteriorDataDistrGauss}
    \pi_D^{y}(y) \propto
    \exp \left[
        -\frac12 \left(y - \tilde{y}\right)^T
        \tilde{C}_D^{-1} 
        \left(y - \tilde{y}\right)
    \right]\, ,
\end{equation}
where the mean and the covariance are given in
Eqs.~\ref{eq:PosteriorDataParamsMean} and~\ref{eq:PosteriorDataParamsCov}. We
can rewrite those expressions as
\begin{align}
    \tilde{y} &= \mathcal{G} \tilde{u} \, , \\
    \tilde{C}_D^{-1} &=
        C_D^{-1} + \left(\mathcal{G} C_M \mathcal{G}^T\right)^{-1}\, .
\end{align}
In order to simplify the notation we introduce
\begin{equation}
    \hat{C}_M = \left(\mathcal{G} C_M \mathcal{G}^T\right)\, ,
\end{equation}
and then
\begin{equation}
    \tilde{C}_D = \hat{C}_M
    - \hat{C}_M \left(\hat{C}_M + C_D \right)^{-1} 
    \hat{C}_M\, .
\end{equation}

%% file: app_datasets.tex
\section{Closure test setup details}
\label{sec:appendix-datasets}

\subsection{Data}

The full list of datasets included in the test set are shown in
Tab.~\ref{tab:summarise_new_data}. The central values are not actually used
in the closure test, however we use the experimental uncertainities in
the calculation of both $\biasvarratio$ and $\xisigdat{1}$. The corresponding
predictions generated from the underlying law are used as the true
observable values. Neither of the data-space closure estimators rely on
the central values of the test datasets.

\begin{table}[h!]
    \begin{center}
        \input{tables/reference_test_data}
    \end{center}
    \caption{
        Observables included in the test data. We wish to stress that the observable
        central values themselves are not used, however the experimental
        uncertainties are used in the definition of the closure estimators, and
        the corresponding predictions from either the underlying law or the
        closure fits.
    }
    \label{tab:summarise_new_data}
\end{table}

\begin{figure}
    \centering
    \includegraphics[width=0.8 \textwidth]{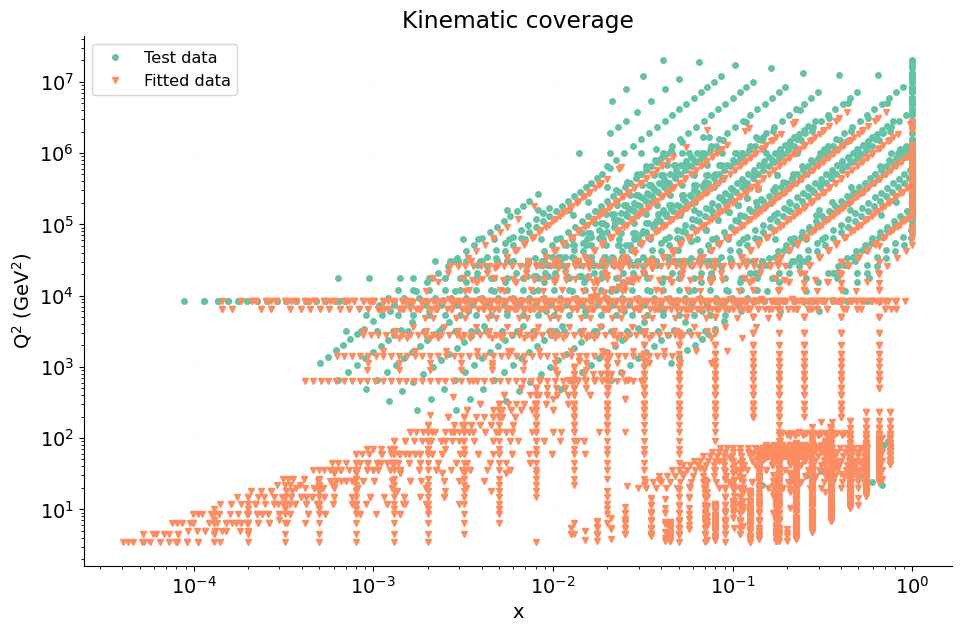}
    \caption{The kinematic coverage of the training and test data
    used to train the models and produce results presented in this paper. 
    The kinematics of the two
    sets of data with this particular split overlaps but there are also kinematic
    regions which the test dataset probes, for which there are no training data.}
    \label{fig:DataKinematicCoverage}
\end{figure}

For completeness, in Fig.~\ref{fig:DataKinematicCoverage}
the kinematic coverage of the training datasets, which
as mentioned is the NNPDF3.1-like dataset used in \cite{Faura_2020}, and the
test datasets shown in Tab.~\ref{tab:summarise_new_data} is plotted.

\subsection{Models}

A summary of the hyperparameters for the neural networks used during the closure
fits is given in table Tab.~\ref{tab:Hyperparams}. These hyperparameters were
chosen as part of an extensive hyperparameters scan, which will be explained in
detail in NNPDF4.0, for this study we simply provide the values of the
hyperparameters as a point of reference.

\begin{table}
    \begin{center}
        \begin{tabular}[h]{c|c}
            \toprule
            hyperparameter & value \\
            \midrule
            architecture & 2-25-20-8 \\
            activation & tanh-tanh-linear \\
            minimiser & NAdam \\
            max training length & 17000 epochs \\
            pre-processing exponents & not fitted \\
            \bottomrule
        \end{tabular}
    \end{center}
    \caption{Hyperparameters for neural networks used in this study. The parameter
    choices, and how these choices were made are discussed in Ref.~\cite{NNPDF40}.}
    \label{tab:Hyperparams}
\end{table}

%% file: tables/reference_test_data.tex
\begin{tabular}{lc}
    \toprule
    Data set
    & Ref.
    \\
    \midrule
    DY E906 $\sigma^d_{\rm DY}/\sigma^p_{\rm DY}$ (SeaQuest)
    & \cite{Dove_2021}
    \\
    \midrule
    ATLAS $W,Z$ 7 TeV ($\mathcal{L}=4.6$~fb$^{-1}$)
    & \cite{Aaboud:2016btc}
    \\
    ATLAS DY 2D 8 TeV
    & \cite{Aaboud:2017ffb}
    \\
    ATLAS high-mass DY 2D 8 TeV
    & \cite{Aad:2016zzw}
    \\
    ATLAS $\sigma_{W,Z}$ 13 TeV
    & \cite{Aad:2016naf}
    \\
    ATLAS $W^+$+jet 8 TeV
    & \cite{Aaboud:2017soa}
    \\
    ATLAS $\sigma_{tt}^{\rm tot}$ 13 TeV ($\mathcal{L} = \SI{139}{\per\femto\barn}$)
    & \cite{Aad:2020tmz}
    \\
    ATLAS $t\bar{t}$ lepton+jets 8 TeV
    & \cite{Aad:2015mbv}
    \\
    ATLAS $t\bar{t}$ dilepton 8 TeV
    & \cite{Aaboud:2016iot}
    \\
    ATLAS single-inclusive jets 8 TeV, R=0.6
    & \cite{Aaboud:2017dvo}
    \\
    ATLAS dijets 7 TeV, R=0.6
    & \cite{Aad:2013tea}
    \\
    ATLAS direct photon production 13 TeV
    & \cite{Aaboud:2017cbm}
    \\
    ATLAS single top $R_{t}$ 7, 8, 13 TeV
    & \cite{Aad:2014fwa,Aaboud:2016ymp,Aaboud:2017pdi}
    \\
    \midrule
    CMS dijets 7 TeV
    & \cite{Chatrchyan:2012bja}
    \\
    CMS 3D dijets 8 TeV
    & \cite{Sirunyan:2017skj}
    \\
    CMS $\sigma_{tt}^{\rm tot}$ 5 TeV
    & \cite{Sirunyan:2017ule}
    \\
    CMS $t\bar{t}$ 2D dilepton 8 TeV
    & \cite{Sirunyan:2017azo}
    \\
    CMS $t\bar{t}$ lepton+jet 13 TeV
    & \cite{Sirunyan:2018wem}
    \\
    CMS $t\bar{t}$ dilepton 13 TeV
    & \cite{Sirunyan:2018ucr}
    \\
    CMS single top $\sigma_{t}+\sigma_{\bar{t}}$ 7 TeV
    & \cite{Chatrchyan:2012ep}
    \\
    CMS single top $R_{t}$ 8, 13 TeV
    & \cite{Khachatryan:2014iya,Sirunyan:2016cdg}
    \\
    \midrule
    LHCb $Z\to \mu\mu, ee$ 13 TeV
    & \cite{Aaij:2016mgv}
    \\
    \bottomrule
\end{tabular}

%% file: app_deltachi2.tex
\section{Understanding NNPDF3.0 data estimators}

In the closure test presented in NNPDF3.0 \cite{nnpdf30} there was a data-space
estimator which aimed to measure the level of over or under fitting, $\deltachi$.
Here we discuss how $\deltachi$ can emerge from the bias-variance decomposition
and then use the linear model to try and understand it in the context of
viewing the ensemble of model replicas as a sample from the posterior distribution
of the model given the data.

Despite the link between the estimators emerging from the decomposition of
$\eout$ and the posterior distribution for data which is not used to inform
the model parameters, if we perform the same decomposition as in
Sec.~\ref{sec:ClosureEstimatorsDerivation} but set
$\testset{\obspriorcent}=\obspriorcent$ then we find that the cross term
in the final line of Eq.~\ref{eq:EoutDecomposition} does not go to zero when
the expectation across data is taken because there is a dependence on
$\obspriorcent$ in both the model predictions and the noisey data. As a result
we have to modify Eq.~\ref{eq:ExpectedBiasVariance} to be
\begin{equation}\label{eq:ExpectedBiasVarianceTraining}
    \mathbf{E}_{\obspriorcent}[\ein] =
    \mathbf{E}_{\obspriorcent}[\bias] + 
    \mathbf{E}_{\obspriorcent}[\var] +
    \mathbf{E}_{\obspriorcent}[{\rm noise}] +
    \mathbf{E}_{\obspriorcent}[\noisecross]\, ,
\end{equation}
where we refer now to the right hand side of
Eq.~\ref{eq:ExpectedBiasVarianceTraining} as $\ein$ because it's evaluated on the
data used to inform the model replicas.

Now we examine the definition of $\deltachi$ introduced
in~\cite{nnpdf30}, defined as the difference between the
$\chi^2$ between the expectation value of the model predictions and the level
one data, and the $\chi^2$ between the underlying observable values and the
level one data. In~\cite{nnpdf30} the denominator was also set to be the
second term in the numerator, however here we slightly re-define
$\deltachi$ to instead simply be normalised by the number of data points:
\begin{equation}\label{eq:deltachi2def1}
    \begin{split}
        \deltachi &= \\
            \frac{1}{\ndata} & \Big[ \left( \emodel{\fwdobsop\left(\modelvecrep\right)} - \obspriorcent \right)^T
            \obspriorcov^{-1}
            \left( \emodel{\fwdobsop\left(\modelvecrep\right)} - \obspriorcent \right) \\
            & \, - \left( \law - \obspriorcent \right)^T
            \obspriorcov^{-1}
            \left( \law - \obspriorcent \right)
        \Big] \\
        &= \bias + \noisecross \, ,
    \end{split}
\end{equation}
where in the second line we show how $\deltachi$ itself can be decomposed to
be equal to two of the terms in Eq.~\ref{eq:ExpectedBiasVarianceTraining}.

Constant values of $\deltachi$ define elliptical contours in data space
centered on the level one data. $\deltachi = 0$, in particular, defines a
contour which is centered on the level one data and passes through the
underlying law. When viewing $\deltachi$ from a classical fitting perspective,
if $\deltachi < 0$ then the expectation value of the model
predictions fit the level one data better than the underlying observables -
which indicates an overfitting of the shift, $\boldsymbol{\shift}$. Similarly,
$\deltachi > 0$ indicates some underfitting of the level one data.

If we return to the linear model we can write the analytic value of
$\deltachi$. Firstly, since $\testset{\obspriorcent}=\obspriorcent$ we can
simplify Eq.~\ref{eq:BiasLinearModel}
\begin{equation}\label{eq:BiasLinearModelSimple}
    \begin{split}
        \mathbf{E}_{\obspriorcent}[{\rm bias}] &= \frac{1}{\ndata}
            {\rm Tr} \left[
                \linmap \modelpostcov \linmap^T \obspriorcov^{-1}
            \right] \\
            &= \frac{1}{\ndata}{\rm Tr} \left[ \modelpostcov \modelpostcov^{-1}\right] \\
            &= \frac{\nmodel}{\ndata} \, ,
    \end{split}
\end{equation}
because $\modelpostcov \modelpostcov^{-1}$ is an $\nmodel \times \nmodel$ identity
matrix. Similarly we can write down the cross term
\begin{equation}
    \begin{split}
        \mathbf{E}_{\obspriorcent}[\noisecross] &= \frac{-2}{\ndata} \mathbf{E}_{\obspriorcent} \left[
            (\linmap \modelpostcov \linmap^T \obspriorcov^{-1} \obsnoise)^T \obspriorcov^{-1} \obsnoise
        \right] \\
        &= \frac{-2}{\ndata}{\rm Tr} [\linmap \modelpostcov \linmap^T \obspriorcov^{-1}] \\
        &= -2 \frac{\nmodel}{\ndata}
    \end{split}
\end{equation}
which leaves us with
\begin{equation}
    \mathbf{E}_{\obspriorcent}[\deltachi] = - \frac{\nmodel}{\ndata} \, .
\end{equation}
The point is that the linear model has already been shown to be a sample from
posterior distribution of the model given the data. But from the classical
fitting point of view we would say this model has overfitted.

As such, we do not report any results with $\deltachi$ here, because
when $\biasvarratio = 1$, it doesn't add much to the discussion. It may still
be useful as a diagnostic tool when $\biasvarratio \neq 1$, which as discussed
could be for a variety of reasons - including fitting inefficiency. It also
may be used as a performance indicator for deciding between two fitting
methodologies: if both fits are shown to have $\biasvarratio = 1$, the
methodology with smaller magnitude of $\deltachi$ could be preferential.
The same could be said for bias and variance however, bias in particular is
clearly closely related to $\deltachi$.